\documentclass[prb,twocolumn,aps,showpacs,eqsecnum,amsmath]{revtex4}
\usepackage{graphicx,bm,times,amsmath,amssymb}
\usepackage{mathrsfs}
\begin{document}
 
\title{Two-loop renormalization-group theory for 
  the quasi-one-dimensional Hubbard model at half filling}

\author{M.\ Tsuchiizu}

\affiliation{
Department of Physics, Nagoya University, Nagoya 464-8602, Japan
}
 
\date{\today}

\begin{abstract}
We derive two-loop renormalization-group equations 
 for the half-filled one-dimensional Hubbard chains
  coupled by the interchain hopping.
Our renormalization-group scheme for the quasi-one-dimensional 
  electron system is a natural extension of that for the 
   purely one-dimensional systems in the sense that
   transverse-momentum dependences are introduced in the 
   $g$-ological coupling constants and
   we regard the transverse momentum as a patch index.
We develop symmetry arguments for
  the particle-hole symmetric half-filled Hubbard model
  and obtain constraints on the $g$-ological coupling constants
  by which resultant renormalization equations are given in 
  a compact form.
By solving the renormalization-group equations numerically, 
  we estimate the magnitude of excitation gaps 
 and clarify that the charge gap 
  is suppressed due to the  interchain hopping  
  but is always finite even for the relevant interchain hopping.
To show the validity of the present analysis, we also apply this 
   to the two-leg ladder system.
By utilizing the field-theoretical bosonization and fermionization method,
  we derive low-energy effective theory 
  and analyze the magnitude of all the excitation gaps in detail. 
It is shown that the low-energy excitations in the two-leg Hubbard ladder
  have  SO(3)$\times$SO(3)$\times$U(1) symmetry 
  when the interchain hopping  exceeds the magnitude of the charge gap.
\end{abstract}

  
\maketitle

\section{Introduction}
 
Renormalization-group (RG) method is one of the 
  most powerful and promising tools to tackle 
  low-dimensional electron and spin systems.
  \cite{Bourbonnais2003}
It has a long history of research
  especially on one dimensional (1D) systems, since
 the RG theory is superior to take into account
  low-dimensional competing fluctuation effects, i.e., 
 it can sum up systematically 
 the logarithmic-singular particle-particle and particle-hole 
  channels which appear in all order of perturbation theory.
 \cite{Emery,Solyom,Bourbonnais1991}
It has been clarified that the RG method describes
    various 1D ground states:
 the Tomonaga-Luttinger (TL) liquid state, 
  the charge-gapped Mott insulating state at half
  filling,  and also the spin-gapped  Luther-Emery state. 
 \cite{Bourbonnais2003,Giamarchi_book}
Not only for the most divergent terms, 
 the next-to-leading logarithmic singular terms 
  have also been studied based on the 
   two-loop formulation of the RG theory,
  \cite{Solyom,Bourbonnais1991}
  where singular self-energy corrections in addition to the 
  vertex corrections are taken into account.
Recently the RG theory 
  is generalized to apply to two-dimensional electron
  systems. \cite{Shankar}
The main difficulty in the RG formulation for two-dimensional systems
  resides in the fact that
   the momentum dependence of the coupling constants 
  is essential but
   the number of independent coupling constants is large and
  it becomes hard to analyze the RG equations even for the one-loop level.
Several attempts have been made 
  by focusing only on dominant 
  scattering processes on the Fermi surface \cite{Furukawa}
  and by discretizing the Fermi surface 
  into finite number of pieces, i.e., so-called patches.
  \cite{Zanchi1998}
For electron systems in arbitrary dimension,
  a nonperturbative RG theory has also been formulated  
  \cite{Salmhofer1998} 
  and has been applied to two dimensional electron systems 
  by considering leading two particle interactions, i.e., 
  within the one-loop level. \cite{Halboth2000,Honerkamp2001}
Quite recently the effect of the two-loop self-energy corrections
  have been examined,
  \cite{Zanchi2001,Honerkamp2003,Katanin2004,Metzner}  while
  the two-loop vertex corrections are considered  only for
  the system with flat Fermi surface.
  \cite{Kishine1999,Freire}

In quasi-one-dimensional (Q1D) electron systems,
the important issue to be clarified is the dimensional
crossover from one to higher dimensions which would 
  occur by changing parameters or temperature. 
  \cite{Bourbonnais2003,Giamarchi_book,Bourbonnais_review}
In real Q1D compounds,
  the TL-liquid behavior is expected at high temperature, 
  however, the effect of warping of the Fermi surface
  due to the small but finite interchain hopping is enhanced 
  at low temperature where the Fermi-liquid behavior can be expected
  if the system is metallic,
  and finally the system has an instability to symmetry-broken states.  
The RG approach is also powerful and succeeds
   in the description of these physical pictures.
  \cite{Bourbonnais2003,Giamarchi_book,Bourbonnais1991,Kishine1998}
In the early RG analysis, the effect of the one-particle interchain hopping
   is treated perturbatively,
however, it is found to be relevant 
  even in the noninteracting case and
 the perturbative treatment is invalid at low temperature.
In order to clarify the dimensional crossover phenomena 
  properly, one has to formulate the RG with the nonperturbative
  treatment of the interchain hopping, i.e., based on 
  the warped Fermi surface.
In this sense, the formulation is analogous to that in 
  the two-dimensional RG scheme since 
  one has to discretize the Fermi surface.
In the Q1D case, the RG has been formulated
  by considering finite number of chains $N_\perp$ 
 ($N_\perp$-chain RG scheme)
 \cite{Lin1997,Duprat2001,Doucot2003,%
   Bourbonnais2004,Fuseya2005,Dupuis2005,Fuseya2006}
 where the transverse momentum 
  is regarded as a patch index.
Based on this scheme, the Q1D systems have been analyzed intensively
  within the one-loop level \cite{Lin1997,Duprat2001,%
        Bourbonnais2004,Fuseya2005,Dupuis2005,Fuseya2006}
  and the self-energy corrections 
  have also been investigated. \cite{Doucot2003}
At commensurate band filling, 
  the dimensional crossover problem becomes nontrivial
  since the 
  electronic correlation has the strongest effect and leads to 
  the Mott insulating state
  if the system is half filled.
The effect of the umklapp scattering between electrons, which 
  is a trigger for the 1D Mott insulator, 
  has been investigated by the one-loop RG,
  \cite{Bourbonnais2004}
 however,  in order to clarify the electronic states 
  in the Mott insulator  one has to 
  examine the properties of the one-particle Green's function,
  i.e., the self-energy corrections,  
  whose singular contributions only appear beyond the one-loop level.
The effects of the two-loop self-energy corrections 
  have also been examined in the Q1D systems
  without considering two-loop vertex corrections,
  \cite{Doucot2003} however,
  a systematic two-loop RG 
  including both the two-loop vertex and self-energy  corrections 
  is not formulated yet.
Recently this issue has also been addressed
  by a numerical method 
  expanding the dynamical mean-field approach 
  (chain-DMFT) \cite{Giamarchi2001,Giamarchi2002,Giamarchi2004,Berthod}  
  and by a field-theoretical method with
  the RPA treatment of the interchain hopping.
  \cite{Essler2002}

From a technical point of view, it is generally hard
  to gain physical insights of results of scaling flows in the Q1D RG,
  since the number of independent coupling constants becomes large
  as $N_\perp$ increases.
As a minimal system of the coupled chains,
  one can consider a two-leg ladder system ($N_\perp=2$).
The two-leg ladder system itself has nontrivial and interesting  
  features \cite{Dagotto} and has been examined intensively
  by using the RG method
  \cite{Fabrizio1993,Nersesyan1993,Khveshchenko1994,Schulz1996,%
  Balents1996,Lin1998,Tsuchiizu1999,Tsuchiizu2002,Fradkin2003,Tsuchiizu2005}
  and also by the high-accuracy numerical 
  technique called the density-matrix-renormalization-group 
  (DMRG) method, \cite{Noack,Weihong}
  where it has been confirmed that
  both the charge and spin modes have excitation gaps
  for the half-filled Hubbard ladder.
In this analysis, one can easily see that
  a naive one-loop RG analysis of the excitation gaps
  is not satisfactory since the RG method 
  breaks down at a energy scale 
  corresponding to the largest excitation gap in a system.
In order to analyze the lower-energy properties,
  one has to derive an effective theory by 
  tracing out the gapped modes
  based on the field-theoretical bosonization/fermionization treatment.
As for the two-leg Hubbard ladder,
  Lin, Balents and Fisher\cite{Lin1998}
   obtained the SO(8) Gross-Neveu model  as 
  an effective theory in the low-energy limit
  and examined the excitation spectrum.
The extended two-leg Hubbard model including additional interactions
  is also examined   \cite{Tsuchiizu2002,Fradkin2003,Tsuchiizu2005} and
  quantum phase transitions between competing ground states have been
  clarified in this context of the one-loop RG.
Despite that the analysis of the two-leg ladder systems
  based on the one-loop RG succeeds in describing the ground state 
  properties, 
  it is not easy to extend the analysis 
  to the case with large number of chains, 
  since the field-theoretical approach is 
  restricted to the small number of chains.
In order to overcome this problem, we formulate, in the present paper,
  the two-loop RG theory for the Q1D electron systems.
Even in the two-loop level, the perturbative approach 
  also breaks down at energy scales of the excitation gaps,
  however,  the respective excitation gaps can be estimated
  by analyzing the scaling behavior
  of the couplings for respective modes,
  without following the tracing-out procedure.
We confirm that the present scheme works 
  even if the respective modes are not independent 
  by revisiting the two-leg ladder systems.

This paper is organized as follows.
In Sec.\ \ref{sec:model}, we introduce
 the finite $N_\perp$-chain half-filled Hubbard model
   coupled by the one-particle interchain hopping, 
 and derive the corresponding  $g$-ology model 
  by linearizing the energy dispersion where 
  the effect of the interchain hopping is treated nonperturbatively.
By developing  symmetry arguments for
  the particle-hole symmetric half-filled Hubbard model, 
  we obtain constraints on the $g$-ological coupling constants. 
In Sec.\ \ref{sec:formulation},
  we formulate the RG based on the Kadanoff-Wilson approach up to 
  the two-loop level, where vertex corrections are taken into account
  based on the third-order perturbation theory, in addition to 
  the second-order calculation for the self-energy corrections.
Reflecting the symmetries that the particle-hole symmetric Hubbard model has, 
  the resultant RG equations can be
  written in a compact form where
  the physical picture can easily be captured.
By solving the RG equations numerically, 
  we estimate the magnitude of the charge and spin gaps.
In Sec.\ \ref{sec:ladder},
  in order to indicate the validity of the present method,
  we consider a most simple but nontrivial 
  case $N_\perp=2$, which corresponds 
  to the two-leg ladder, and analyze the 
  excitation properties in detail 
  by combining the field-theoretical bosonization and  
  fermionization method.
Finally, the results are summarized in Sec.\ \ref{sec:summary}.
Technical details are given in the Appendices A and B.
In the Appendix C, we give a related issue which supports strongly 
  the validity of the present estimation of excitation gaps.

\section{Model and symmetry arguments}\label{sec:model}

We consider the bipartite Q1D Hubbard model at half filling
 with $t_\parallel \gg t_\perp$, where
  the transfer integral along chains is $t_\parallel$ 
  and that between chains is $t_\perp$.
Our Hamiltonian is given by
%====================================================================
\begin{eqnarray}
H &=& 
-t_\parallel \sum_{j,l,s} 
\left(c_{j,l,s}^\dagger c_{j+1,l,s}+\mathrm{H.c.} \right)
\nonumber \\ && {}
-t_\perp \sum_{j,l,s} 
\left(c_{j,l,s}^\dagger c_{j,l+1,s}+\mathrm{H.c.} \right)
\nonumber \\ && {}
+ U \sum_{j,l} n_{j,l,\uparrow} n_{j,l,\downarrow} ,
\label{eq:model}
\end{eqnarray}
%====================================================================
where $c_{j,l,s}$ is the annihilation operator of electron 
  on the $j$th site in the $l$th chain with spin $s$, and 
 $n_{j,l,s}=c_{j,l,s}^\dagger c_{j,l,s}^{}-\frac{1}{2}$.
The system size along chains ($N_\parallel$) 
  is considered to be sufficiently large
  and the sum of the site index, which runs $j=1,\cdots,N_\parallel$, 
 is to be understood as an integral 
  in the thermodynamic limit.
The chain index runs $l=1,\cdots,N_\perp$ and
we consider the system with  finite number of chains $N_\perp$ where
 the periodic boundary condition is imposed
 $c_{j,N_\perp+1,s}=c_{j,1,s}$.

\subsection{$g$-ology notation}

%====================================================================
\begin{figure}[b]
\includegraphics[width=4cm]{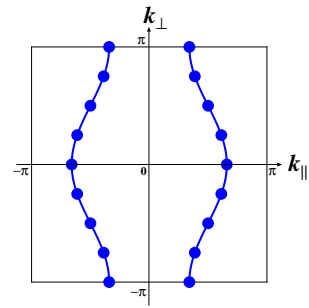}
\caption{
(Color online)
Fermi points (closed circles) 
  in the present half-filled Q1D Hubbard model with
  the periodic boundary condition in the
  transverse direction.
The case for $N_\perp=8$ is shown.
}
\label{fig:fs}
\end{figure}
%======================================================================

The kinetic term of the Hamiltonian is given by
%====================================================================
\begin{eqnarray}
H_0 
&=&
\sum_{\bm k,s}
\varepsilon(\bm k) \,
c_s^\dagger (\bm k) \, c_s^{}(\bm k) ,
\\
\varepsilon(\bm k) &=& -2t_\parallel \cos k_\parallel 
-2t_\perp \cos k_\perp ,
\end{eqnarray}
%====================================================================
where $\bm k = (k_\parallel,k_\perp)$ and
the lattice constant is set to unity.
Since the system is particle-hole symmetric,
  we can assume $t_\parallel > 0$ and $t_\perp \ge 0$
  without losing generality.
Since the number of chains $N_\perp$ is finite,
  the transverse momentum is given by
%====================================================================
\begin{eqnarray}
k_\perp =\frac{2\pi}{N_\perp}n, \quad
 n=-\left[\frac{N_\perp}{2}\right], ... , \left[\frac{N_\perp}{2}\right],
\label{eq:kperp}
\end{eqnarray}
%====================================================================
where $[x]$ is the Gauss symbol 
 denoting maximum integer which does not exceed $x$.
By assuming $t_\perp \ll t_\parallel$,
  we linearize the dispersion where the situation can be 
  simplified as follows.
Up to the lowest order in $t_\perp$
 the kinetic term with the linearized dispersion is given by
%====================================================================
\begin{eqnarray}
H_0 
&=&
\sum_{\bm k,p,s}
\varepsilon_p(\bm k) \,
c_{p,s}^\dagger (\bm k) \, c_{p,s}^{}(\bm k) ,
\label{eq:kinetic}
\\
\varepsilon_p (\bm k)
 &=&  v(pk_\parallel -k_F)
-2t_\perp \cos k_\perp,
\label{eq:lineardisp}
\end{eqnarray}
%====================================================================
where $v=2t_\parallel$ and $k_F=\pi/2$.
We introduce the bandwidth cutoff $\Lambda$.
In this approximation, the warped open Fermi surface (Fig.\ \ref{fig:fs})
  is specified as a function
  of $k_\perp$:
%====================================================================
\begin{equation}
k_F(k_\perp)  =  
k_F +2 \frac{t_\perp}{v} \cos k_\perp ,
\label{eq:kf}
\end{equation}
%====================================================================
and the energy dispersion (\ref{eq:lineardisp}) can be reexpressed
   as $\varepsilon_p (\bm k) =  v[pk_\parallel -k_F(k_\perp)]$.
Thus we regard the transverse momentum $k_\perp$
  as a \textit{patch index} in the present RG formulation.
The greatest merit of the present formulation lies in the fact that 
   the transverse momentum $k_\perp$ is a conserved quantity, i.e.,
  the patch index is a good quantum number and 
  the ambiguity of selecting  patch index disappears.

%====================================================================
\begin{figure*}[t]
\includegraphics[width=15cm]{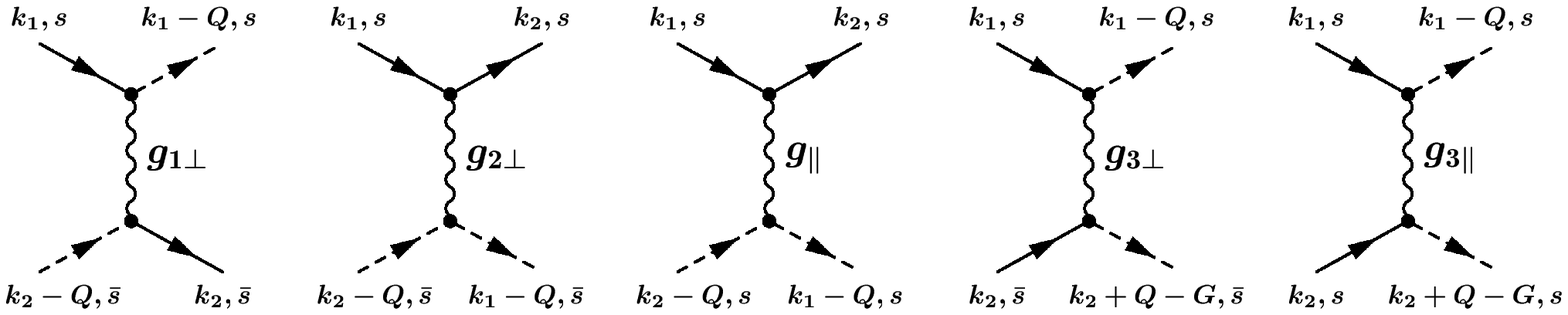}
\caption{
$g$-ology notation. The solid (dashed) line 
 denotes a right-moving (left-moving) electron.  
$\bm k_i = (k_{\parallel i}, k_{\perp i})$,
$\bm Q=(\pi+q_\parallel,q_\perp)$, and 
$\bm G =(2\pi, 0)$. 
}
\label{fig:gology}
\end{figure*}
%======================================================================

Following the conventional $g$-ology approach, \cite{Solyom}
 we classify the interaction part of the 
  Hamiltonian $H_{\mathrm{I}}=U\sum_{j,l} n_{j,l,\uparrow} n_{j,l,\downarrow}$
  into the forward, backward, and umklapp scattering processes,
  by focusing on the longitudinal momentum $k_\parallel$.
We introduce the coupling constants
  $g_{1\perp}$, $g_{2\perp}$,
  $g_\parallel$, $g_{3\perp}$, and $g_{3\parallel}$,
  which represent
   the backward scattering with the opposite spins ($g_{1\perp}$),
   the forward scattering with the opposite spins ($g_{2\perp}$),
   the forward scattering with the same spins ($g_\parallel$),
   the umklapp  scattering with the opposite spins ($g_{3\perp}$),
  and the umklapp  scattering with the same spins ($g_{3\parallel}$).
In terms of the Hubbard interaction $U$, 
  the magnitudes of the couplings are
  given by   $g_{1\perp}=g_{2\perp}=g_{3\perp}=U$ and 
  $g_\parallel=g_{3\parallel}=0$.
The $g_\parallel$ and $g_{3\parallel}$ processes are absent
  in the original Hubbard
 interactions, however, can become finite under the RG scaling procedure. 
Furthermore,  the coupling constants are 
  differently renormalized depending on the external transverse momenta 
  of the vertex 
   and have the explicit transverse-momentum (i.e., patch-index) dependence.
To take into account these effects, we formally introduce 
  the transverse momentum dependence of the coupling constants 
  in the initial $g$-ology Hamiltonian.
In the most general form,
  the interaction part of the Hamiltonian is given by
\begin{widetext}
%====================================================================
\begin{eqnarray}
H_\mathrm{I} &=& 
+\frac{1}{V}\sum_{\bm k_1,\bm k_2, \bm q,s}
  g_{1\perp(q_\perp,k_{\perp 1},k_{\perp 2})} \,\,
 c_{+,s}^\dagger(\bm k_1) \,
 c_{-,s}^{} (\bm k_1- \bm Q) \,\,
 c_{-,\bar s}^\dagger (\bm k_2-\bm Q) \,
 c_{+,\bar s}^{}(\bm k_2)
\nonumber \\ &&{}
+\frac{1}{V}\sum_{\bm k_1,\bm k_2, \bm q ,s} 
 g_{2\perp (q_\perp,k_{\perp 1},k_{\perp 2})} \,\,
 c_{+,s}^\dagger (\bm k_1)  \,
 c_{+,s}^{}(\bm k_2) \,\,
 c_{-,\bar s}^\dagger (\bm k_2- \bm Q) \,
 c_{-,\bar s}^{}(\bm k_1-\bm Q)
\nonumber \\ &&{}
+\frac{1}{V}\sum_{\bm k_1,\bm k_2, \bm q,s} 
 g_{\parallel (q_\perp,k_{\perp 1},k_{\perp 2})}\,\,
 c_{+,s}^\dagger (\bm k_1)  \,
 c_{+,s}^{}(\bm k_2) \,\,
 c_{-,s}^\dagger (\bm k_2- \bm Q) \,
 c_{-,s}^{}(\bm k_1-\bm Q)
\nonumber \\ &&{}
+\frac{1}{2V}\sum_{\bm k_1,\bm k_2, \bm q,s} 
   g_{3\perp (q_\perp,k_{\perp 1},k_{\perp 2})} 
 \Bigl[ 
 c_{+,s}^\dagger (\bm k_1)  \, 
 c_{-,s}^{} (\bm k_1- \bm Q) \,\,
 c_{+,\bar s}^\dagger (\bm k_2)  \, 
 c_{-,\bar s}^{}(\bm k_2+\bm Q - \bm G)
  +\mathrm{H.c.}\Bigr]
\nonumber \\ &&{}
+\frac{1}{2V}\sum_{\bm k_1,\bm k_2, \bm q, s} 
   g_{3\parallel (q_\perp,k_{\perp 1},k_{\perp 2})} \,
 \Bigl[ 
 c_{+,s}^\dagger (\bm k_1)  \,
 c_{-,s}^{} (\bm k_1- \bm Q)  \,\,
 c_{+,s}^\dagger (\bm k_2)  \,
 c_{-,s}^{}(\bm k_2+ \bm Q - \bm G)
  +\mathrm{H.c.}\Bigr],
\label{eq:g-ology}
\end{eqnarray}
%====================================================================
\end{widetext}
where $\bar s=\uparrow$$(\downarrow)$ for $s=\downarrow$$(\uparrow)$,
and $\bm Q=(\pi+q_\parallel,q_\perp)$, 
$\bm G =(2\pi, 0)$, and $V=N_\parallel N_\perp$.
The momenta $k_{\parallel 1}$ and
  $k_{\parallel 2}$ are assumed to take values near  $k_F(k_\perp)$.
In the transverse direction, on the other hand,
 the momenta $k_{\perp i}$ and $q_\perp$
  can take the values in $-\pi<k_{\perp i},q_\perp \le \pi$,
  and the momentum $(k_{\perp i} \pm q_\perp)$ is assumed to 
  reduce the first Brillouin zone,
   then  all the possible scattering processes are taken into account,
  including the transverse umklapp scattering. 
The respective scattering processes are shown in Fig.\ \ref{fig:gology}.
We will neglect the forward scattering with the same branch,
  so-called $g_4$ term, since
  this process does not show the logarithmic-singular behavior  in 
  perturbation and is known to 
  yield only quantitative changes in velocities for the 1D case. 
In terms of the Hubbard interaction $U$, 
  the magnitudes of the couplings are
  given by   
  $g_{1\perp(q_\perp,k_{\perp 1},k_{\perp 2})}
  =g_{2\perp(q_\perp,k_{\perp 1},k_{\perp 2})}
  =g_{3\perp(q_\perp,k_{\perp 1},k_{\perp 2})}=U$ and 
  $g_{\parallel(q_\perp,k_{\perp 1},k_{\perp 2})}
  =g_{3\parallel(q_\perp,k_{\perp 1},k_{\perp 2})}=0$.
To simplify the notation,
  we will suppress the $\perp$ index of the transverse momentum
  in the following.
All the coupling constants are assumed to be real.
In order to make  $H_\mathrm{I}$  hermitian,
   the coupling constants must satisfy 
%====================================================================
\begin{eqnarray}
g_{1\perp(q,k_1,k_2)} &=& g_{1\perp(q,k_2,k_1)}, \nonumber \\
g_{2\perp(q,k_1,k_2)} &=& g_{2\perp(q,k_2,k_1)}, \nonumber \\
g_{\parallel(q,k_1,k_2)} &=& g_{\parallel(q,k_2,k_1)}, 
\label{eq:hermite} \\
g_{3\perp(q,k_1,k_2)} &=& g_{3\perp(-q,k_2,k_1)},\nonumber \\
g_{3\parallel(q,k_1,k_2)} &=& g_{3\parallel(-q,k_2,k_1)}.\nonumber
\end{eqnarray}
%====================================================================
As in the 1D case, the physical picture  becomes transparent by introducing  
 a new set of the couplings:
%====================================================================
\begin{eqnarray}
g_{\rho(q,k_1,k_2)}
&\equiv&
g_{2\perp(q,k_1,k_2)}+g_{\parallel(q,k_1,k_2)},
\nonumber\\ 
g_{\sigma(q,k_1,k_2)}
&\equiv&
g_{2\perp(q,k_1,k_2)}-g_{\parallel(q,k_1,k_2)},
\nonumber\\
g_{c(q,k_1,k_2)}
&\equiv&
g_{3\perp(q,k_1,\pi-k_2)},
\label{eq:notation}\\
g_{s(q,k_1,k_2)}
&\equiv&
g_{1\perp(q,k_1,k_2)},
\nonumber\\
g_{cs(q,k_1,k_2)}
&\equiv&
g_{3\parallel(q,k_1,\pi-k_2)},
\nonumber
\end{eqnarray}
%====================================================================
where $g_\rho$ and $g_c$ ($g_\sigma$ and $g_s$) are
  the coupling constants representing the charge (spin) degrees of 
  freedom.
This picture can easily be captured by noting 
  that, if we neglect the momentum dependence of the coupling constants,
  the  $g_{1\perp}$, $g_{2\perp}$, $g_\parallel$, and $g_{3\perp}$ terms 
    of the Hamiltonian (\ref{eq:g-ology})
  are written in symmetric forms as\cite{Gogolin}
%====================================================================
\begin{eqnarray}
\hspace*{-1.cm}&& {}
-\frac{g_\sigma}{V}\sum_{p,\bm q}
 J^z_{p}(\bm q) 
 J^z_{-p}(-\bm q) 
-\frac{g_s}{V}\sum_{p,\bm q}
   J^+_{p}(\bm q)  J^-_{-p}(-\bm q)
\nonumber \\ 
\hspace*{-.5cm}&&{}
+\frac{g_\rho}{V}\sum_{p,\bm q}
 J'^z_p(\bm q) 
 J'^z_{-p}(-\bm q) 
+\frac{g_c}{V}\sum_{p,\bm q} 
    J'^+_p(\bm q)  J'^-_{-p}(-\bm q),
\qquad 
\end{eqnarray}
%====================================================================
where the respective chiral density operators are given by
%====================================================================
\begin{subequations}
\begin{eqnarray}
J^z_p(\bm q) &=&\frac{1}{2}
\sum_{\bm k,s} 
 \Bigl[ c_{p,\uparrow}^\dagger(\bm k) \,  c_{p,\uparrow}^{} (\bm k+\bm q)
\nonumber \\ && {} \qquad
-  c_{p,\downarrow}^\dagger(\bm k) \,  c_{p,\downarrow}^{} (\bm k+\bm q)
 \Bigr],
\\
J'^z_p(\bm q) &=&\frac{1}{2}
\sum_{\bm k,s}
  :  c_{p,s}^\dagger(\bm k) \,  c_{p,s}^{} (\bm k+\bm q)   :, 
\\
J^-_p(\bm q) &=&
\sum_{\bm k}
  c_{p,\downarrow}^\dagger(\bm k) \,  c_{p,\uparrow}^{} (\bm k+\bm q),
\\
J'^-_p(\bm q) &=&
\sum_{\bm k}
c_{p,\uparrow} (\bm k) \,  
c_{p,\downarrow} \biglb((\pi,\pi)-\bm k + \bm q\bigrb), \quad
\end{eqnarray}
\end{subequations}
%====================================================================
and  $J^+_p(\bm q)= [J^-_p(-\bm q)]^\dagger$,
   $J'^+_p(\bm q)= [J'^-_p(-\bm q)]^\dagger$.
In the 1D half-filled Hubbard model ($g_\rho=g_c$ and $g_\sigma=g_s$), 
  it is known that
  the charge part, in addition to the spin one, also becomes  SU(2) symmetric.
  \cite{Gogolin}
Even in the Q1D case, the model has an additional SU(2) symmetry, which 
 is shown explicitly in Sec.\ \ref{sec:psedospinSU(2)}.
The $G_{cs}$ coupling represents the spin-charge coupling term in the 1D 
  case  as seen from the bosonization technique.
 \cite{Tsuchiizu_Furusaki} 
In the notation of Eq.\ (\ref{eq:notation}),
   the conditions of the hermitian (\ref{eq:hermite})
  can be expressed as
  $g_{\nu(q,k_1,k_2)}=g_{\nu(q,k_2,k_1)}$,
  for $\nu=\rho,\sigma,s$ and
  $g_{\nu(q,k_1,k_2)}=g_{\nu(-q,\pi-k_2,\pi-k_1)}$ 
   for $\nu=c,cs$.
The number of  independent coupling constants $g_i$ 
in Eq.\ (\ref{eq:g-ology}) is   $5N_\perp^2(N_\perp+1)/2$.

\subsection{Symmetry arguments}

The Hubbard model (\ref{eq:model}) is known to have high 
  symmetries, however, 
  the $g$-ology Hamiltonian
  (\ref{eq:g-ology}) is generalized one including 
  low symmetry.
Reflecting symmetries that the Hubbard model has,
  there appear several constraints on the $g$-ological couplings
  and the resultant RG equations can be simplified.
In this subsection, we clarify relations for the coupling 
  constants protected by the symmetries.

\subsubsection{Spin-rotational SU(2)}

The Hubbard model (\ref{eq:model}) 
  is invariant under spin-rotation, while the $g$-ology Hamiltonian
  (\ref{eq:g-ology}) includes the 
  spin-anisotropic case.
The spin-rotational symmetry can be argued in terms of 
the generators of the spin rotation which are nothing but the spin operator:
%====================================================================
\begin{eqnarray}
\bm S =\frac{1}{2}
 \sum_{\bm k,s_1,s_2}  c_{s_1}^\dagger(\bm k) \, \bm \sigma_{s_1,s_2} 
   c_{s_2}^{} (\bm k).
\label{eq:spinoperator}
\end{eqnarray}
%====================================================================
The arbitrary global spin rotation
  by these generators can be represented by the SU(2) matrix:
%====================================================================
\begin{eqnarray}
\left(
\begin{array}{c}
c_{j,l,\uparrow} \\ c_{j,l,\downarrow}
\end{array}
\right)
 \to
\left(
\begin{array}{cc}
  a_\sigma   & b_\sigma \\
- b_\sigma^* & a_\sigma^* 
\end{array}
\right)
\left(
\begin{array}{c}
c_{j,l,\uparrow} \\ c_{j,l,\downarrow}
\end{array}
\right),
\label{eq:spinSU(2)}
\end{eqnarray}
%====================================================================
where $a_\sigma$ and $b_\sigma$ are complex numbers satisfying
   $|a_\sigma|^2+|b_\sigma|^2=1$.
Obviously the Hubbard Hamiltonian (\ref{eq:model}) is invariant
  under the transformation (\ref{eq:spinSU(2)}).
By requiring the $g$-ology Hamiltonian 
  (\ref{eq:g-ology}) to be invariant under this rotation,
we obtain the constraints on the coupling constants.
In the notation (\ref{eq:notation}), the constraint relations 
  are given by
%====================================================================
\begin{subequations}
\begin{eqnarray}
&&
g_{s(q,k_1,k_2)}
=
g_{\sigma(q,k_1,k_2)},
\label{eq:spinSU(2)gs_gsigma}
\\
&&
g_{c(q,k_1,k_2)}-
g_{c(\pi-q+k_1+k_2,k_1,k_2)}
\nonumber \\
&& \qquad
=
g_{cs(q,k_1,k_2)}-
g_{cs(\pi-q+k_1+k_2,k_1,k_2)}. 
\label{eq:spinSU(2)gs_gcs}
\qquad\qquad
\end{eqnarray}%
\label{eq:spinSU(2)_gology}%
\end{subequations}
%====================================================================
Since the property of the spin-rotational invariance
   is hold under the RG procedure, 
 these relations can be considered as 
   constraints on the renormalized   coupling constants.

\subsubsection{Particle-hole symmetry}

The present bipartite half-filled system is invariant under
the particle-hole transformation 
  $c_{s}(\bm k)\leftrightarrow c_{s}^\dagger 
 \biglb((\pi,\pi)-\bm k\bigrb)$,
 where $c_s(\bm k)$ is the Fourier transform of $c_{j,l,s}$.
In the linearized dispersion (\ref{eq:lineardisp}),
  this particle-hole transformation corresponds to 
%====================================================================
\begin{eqnarray}
c_{p,s}(\bm k)\leftrightarrow c_{p,s}^\dagger 
 \biglb((p\pi,\pi)-\bm k\bigrb).
\end{eqnarray}
%====================================================================
In order to make this particle-hole symmetry meaningful,
  the number of chains $N_\perp$ must be even, otherwise the
  $k_\perp$ [Eq.\ (\ref{eq:kperp})] cannot become symmetric 
  in this transformation.
By imposing the condition that the $g$-ology Hamiltonian 
  (\ref{eq:g-ology}) is invariant under this rotation,
we obtain the constraints, 
in the notation (\ref{eq:notation}),
%====================================================================
\begin{eqnarray}
g_{\nu(q,k_1,k_2)}
&=& g_{\nu(-q,\pi-k_1,\pi-k_2)}
,
\label{eq:ph} 
\end{eqnarray}
%====================================================================
where $\nu=\rho,\sigma,c,s,cs$.
We note that, by combining the relation 
   $g_{c/cs(q,k_1,k_2)}=g_{c/cs(-q,\pi-k_2,\pi-k_1)}$ 
  [obtained from Eq.\ (\ref{eq:hermite})], 
 we find  $g_{c/cs(q,k_1,k_2)}=g_{c/cs(q,k_2,k_1)}$.

\subsubsection{Pseudospin SU(2)}\label{sec:psedospinSU(2)}

In addition to the particle-hole symmetry,
the system has an additional  symmetry,
if the interaction is on-site one only.\cite{Yang}
The generators of this SU(2) are given by\cite{Yang}
%====================================================================
\begin{subequations}
\begin{eqnarray}
&&
Q^x
\equiv  \frac{\eta^\dagger + \eta}{2},
\quad
Q^y
\equiv \frac{\eta^\dagger-\eta}{2i},
\\
&&
Q^z
\equiv 
\frac{1}{2} \sum_{\bm k,s}  
:c_{s}^\dagger (\bm k) \,  c_{s}^{} (\bm k)  : ,
\end{eqnarray}%
\label{eq:chargeoperator}%
\end{subequations}
%===============================================================
where the so-called $\eta$-pairing operator is given by
%====================================================================
\begin{eqnarray}
\eta
\equiv 
 \sum_{\bm k}  
c_{\uparrow} (\bm k) \,  c_{\downarrow} \biglb((\pi,\pi)-\bm k\bigrb).
\end{eqnarray}
%===============================================================
The arbitrary rotation by these generators can be 
  represented by the SU(2) matrix:
%====================================================================
\begin{eqnarray}
\left(
\begin{array}{c}
 c_{j,l,\uparrow} \\
 c_{j,l,\downarrow}^\dagger
\end{array}
\right)
\to
\left(
\begin{array}{cc}
 a_\rho & z_{j,l} b_\rho \\
- z_{j,l} b_\rho^* & a_\rho^*
\end{array}
\right)
\left(
\begin{array}{c}
c_{j,l,\uparrow} \\
c_{j,l,\downarrow}^\dagger
\end{array}
\right),
\label{eq:chargeSU(2)}
\end{eqnarray}
%====================================================================
where $a_\rho$ and $b_\rho$ are complex numbers satisfying
   $|a_\rho|^2+|b_\rho|^2=1$, and $z_{j,l}=(-1)^{j+l}$.
This transformation commutes with Eq.\ (\ref{eq:spinSU(2)}).
One easily finds that
  this symmetry breaks down if the Hubbard model is extended, e.g.,
  by including an additional intersite interaction.
In the Fourier space with the linearized dispersion,
 the transformation (\ref{eq:chargeSU(2)}) corresponds to 
%====================================================================
\begin{subequations}
\begin{eqnarray}
c_{p,\uparrow}(\bm k)  
&\to&  
  a_\rho c_{p,\uparrow}  (\bm k) 
 + b_\rho c_{p,\downarrow}^\dagger \biglb( (p\pi,\pi)-\bm k \bigrb),
\\
c_{p,\downarrow}^\dagger (\bm k) 
&\to&
a_\rho^* c_{p,\downarrow}^\dagger  (\bm k) 
-b_\rho^* c_{p,\uparrow}  \biglb((p\pi,\pi)-\bm k \bigrb). \quad
\end{eqnarray}%
\end{subequations}
%====================================================================
The kinetic term (\ref{eq:kinetic}) is invariant 
under this transformation. 
By imposing the condition  that the $g$-ology Hamiltonian
  (\ref{eq:g-ology}) is invariant under this transformation,
  we obtain 
%====================================================================
\begin{subequations}
\begin{eqnarray}
&&
g_{c(q,k_1,k_2)}+g_{c(\pi-q+k_1+k_2,k_1,k_2)}
\nonumber \\
&&{}\qquad =
+ g_{\rho(q,k_1,k_2)}+g_{\rho(\pi-q+k_1+k_2,k_1,k_2)},
\label{eq:chargeSU(2)gc_grho}
\\ 
&&
g_{c(q,k_1,k_2)}-g_{c(\pi-q+k_1+k_2,k_1,k_2)}
\nonumber \\
&&{}\qquad =
 -g_{\sigma(q,k_1,k_2)}+g_{\sigma(\pi-q+k_1+k_2,k_1,k_2)}, \qquad\qquad
\\ 
&&
g_{s(q,k_1,k_2)}-g_{s(\pi-q+k_1+k_2,k_1,k_2)}
\nonumber \\
&&{}\qquad =
 -g_{cs(q,k_1,k_2)}+g_{cs(\pi-q+k_1+k_2,k_1,k_2)}. \qquad\qquad
\label{eq:chargeSU(2)gs_gcs}
\end{eqnarray}%
\label{eq:chargeSU(2)_gology}%
\end{subequations}
%====================================================================
The first relation is a natural extension to the known relation 
  for the purely 1D case.\cite{Gogolin}
The last two relations, which do not appear in the 1D limit,
  imply
  that the couplings $g_c$ and $g_\sigma$ ($g_s$ and $g_{cs}$) are 
  not independent and related to each other.

The relations (\ref{eq:chargeSU(2)_gology}) can also be derived 
  from  the spin SU(2) relations (\ref{eq:spinSU(2)_gology})
  by using the charge-spin duality relation, 
  as explicitly shown in the Appendix \ref{sec:appendix_duality}.

\subsection{Two-loop RG theory for the 1D Hubbard model}

We briefly recall the known results
  of the two-loop RG theory for the purely 1D case,
by focusing on the half-filled 1D Hubbard model:
%====================================================================
\begin{equation}
H_{\mathrm{1D}} =
-t \sum_{j,s} 
\left(c_{j,s}^\dagger c^{}_{j+1,s}+\mathrm{H.c.} \right)
+ U \sum_{j,l} n_{j,\uparrow} n_{j,\downarrow},
\end{equation}
%====================================================================
where $c_{j,s}$ is the annihilation operator of electron 
  on the $j$th site with spin $s$, and 
 $n_{j,s}=c_{j,s}^\dagger c_{j,s}^{}-\frac{1}{2}$. 
The linearized dispersion is
$\varepsilon (k) = -2t\cos k
 \to  v(\pm k_\parallel -k_F) $
where the Fermi velocity and the Fermi momentum are
  $v=2t$ and $k_F=\pi/2$.
The $g$-ological scattering matrices are the same as Fig.\
\ref{fig:gology} and we introduce
$g_{\rho} \equiv (g_{2\perp}+g_{\parallel})$,
$g_{\sigma} \equiv (g_{2\perp}-g_{\parallel})$,
$g_{c} \equiv g_{3\perp}$,
$g_{s} \equiv g_{1\perp}$, and
$g_{cs} \equiv g_{3\parallel}$, as before.
The two-loop RG equations for the respective couplings are given by
\cite{Solyom}
%====================================================================
\begin{subequations}
\begin{eqnarray}
\frac{d}{dl}
G_{\rho}
&=&
+ 2G_c^2 - 2 G_{\rho} G_{c}^2 ,
\\
\frac{d}{dl}
G_{c}
&=&
+2 G_{\rho}  G_{c}  -  G_{\rho}^2  G_{c}  -  G_{c}^3 ,
\\
\frac{d}{dl} G_{\sigma}
&=& 
- 2G_{s}^2- 2 G_{\sigma} G_s^2 ,
\\
\frac{d}{dl} G_{s}
&=& 
- 2G_{\sigma} G_s - G_{\sigma}^2 G_s -  G_{s}^3 ,
\end{eqnarray}%
\label{eq:RG1dbare}%
\end{subequations}
%====================================================================
where $l$ is the scaling parameter and the initial values are
  given by 
  $G_i(0)=g_i/(2\pi v)$.
We have neglected the $G_{cs}$ coupling, 
   since this has an irrelevant canonical dimension.
 \cite{Tsuchiizu_Furusaki} 

These RG equations can be simplified reflecting the symmetries
  of the system.
  The spin-rotational SU(2) symmetry ensures 
  $G_\sigma(l)=G_s(l)$, which is obtained from 
  Eq.\ (\ref{eq:spinSU(2)gs_gsigma})
  by neglecting the transverse momentum dependences.
This relation holds  even under the scaling procedure.
The particle-hole symmetric Hubbard model 
  has another pseudospin SU(2) symmetry 
  and then the total Hamiltonian is characterized by 
  the SU(2)$\times$SU(2) symmetry.\cite{Gogolin}
This pseudospin SU(2) symmetry ensures $G_\rho(l)=G_c(l)$,
   which can be obtained
  from Eq.\ (\ref{eq:chargeSU(2)gc_grho}), and 
  thus this can be considered as the 
  ``charge'' SU(2) symmetry.
In this SU(2)$\times$ SU(2) symmetric case, 
  the RG equations (\ref{eq:RG1dbare}) can be simplified as
%====================================================================
\begin{subequations}
\begin{eqnarray}
\frac{d}{dl}
G_{\rho}
&=&
+ 2G_\rho^2 - 2 G_{\rho}^3,
\label{eq:RG1d_rho}
\\
\frac{d}{dl} G_{\sigma}
&=& 
- 2G_{\sigma}^2- 2 G_{\sigma}^3,
\end{eqnarray}%
\label{eq:RG1d}%
\end{subequations}
%====================================================================
where the initial values are given by 
  $G_\rho(0)=G_\sigma(0)=U/(2\pi v)$.
For repulsive interaction $U>0$, 
  one finds from Eq.\ (\ref{eq:RG1d}) that 
 the $G_\sigma(l)$ coupling decreases under scaling
  and  is  marginally irrelevant,
 while  $G_\rho(l)$ is  marginally relevant.
The relevance/irrelevance of the couplings reflects the 
  low-energy properties having finite/zero excitation
  gap in the corresponding modes.
This behavior correctly reflects the properties of the 1D Mott insulator,
  where only the charge degrees of freedom is frozen due to 
  the finite Mott gap  and the spin has gapless excitations.
By integrating out Eq.\ (\ref{eq:RG1d_rho}) analytically from 
  $l=0$ to $l=l_\rho \equiv\ln(\Lambda/\Delta_\rho)$,
   one can obtain the characteristic energy scale $\Delta_\rho$ as
%====================================================================
\begin{eqnarray}
\Delta_\rho = C_\rho \Lambda \sqrt{G_\rho} \exp(-1/2G_\rho),
\label{eq:1DMottgap}
\end{eqnarray}%
%====================================================================
where $C_\rho$ is an integration constant depending on
$G_\rho(l_\rho)$.
This formula reproduces the exactly known Mott gap in one dimension 
  in the weak $U$ region, since
  the $U$ dependence of $\Delta_\rho$ is given by 
  $\Delta_\rho \propto  \sqrt{t_\parallel U} \exp(-2\pi t_\parallel/U)$.
  \cite{Ovchinikov}

\section{Formulation of renormalization group}\label{sec:formulation}

In this section, we derive the RG equations for  the Q1D half-filled 
  Hubbard model in the two-loop level
  based on the Kadanoff-Wilson cutoff scaling scheme. 
 \cite{Bourbonnais2003}
In the one-loop level, the formulation for the Q1D case is  found in
  Refs.\  \onlinecite{Duprat2001,Bourbonnais2004,Fuseya2005,Dupuis2005,%
    Fuseya2006,Doucot2003}.
In this scheme,
  we take partial integration of the partition function over the fermion
   degrees of freedom in the outer energy shell and
   scale the bandwidth cutoff $\Lambda$  as
   $\Lambda_l= \Lambda e^{-l}$ where $l$ is the scaling parameter.
We perform the logarithmic approximation, i.e., we keep the diagrams
  which become logarithmic singular in the 1D limit and thus
  the resultant Q1D RG equations are natural extensions to those for 
   purely 1D case.
In order to simplify the notations, 
we introduce the dimensionless couplings
  $G_{\nu(q,k_1,k_2)}\equiv g_{\nu(q,k_1,k_2)}/2\pi v$,
  where $\nu=\rho,\sigma,c,s,cs$.

\subsection{Peierls and Cooper bubbles in the one-loop level}

%====================================================================
\begin{figure}[t]
\includegraphics[width=7.5cm]{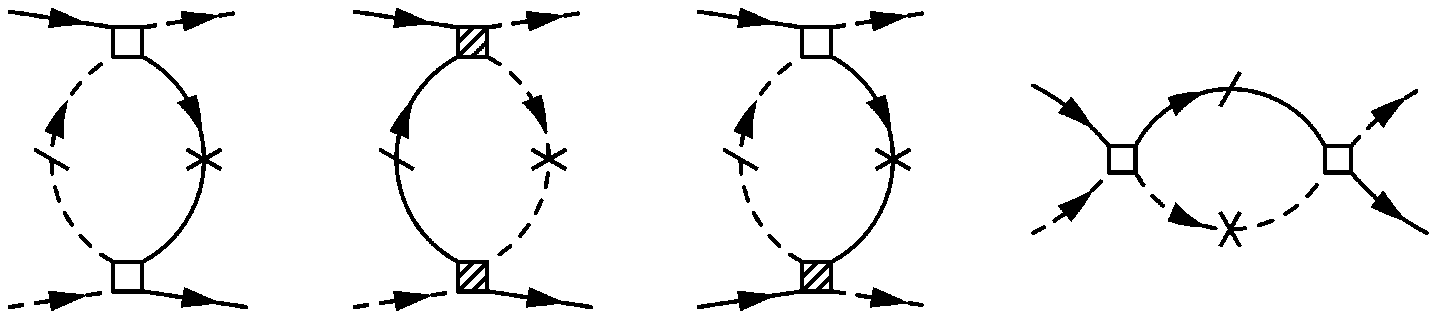}
\caption{
The second-order diagrams contributing to the vertex corrections.
The open square is the vertex for forward and backward scatterings, i.e.,
 $g_{1\perp}$, $g_{2\perp}$, and $g_{\parallel}$, and 
  the shaded square is the one for umklapp scattering $g_{3\perp}$ and
 $g_{3\parallel}$.
The solid (dashed) line refers to a right-moving (left-moving) electron,
  $p=+(-)$.
The slashed line represents that the electron has energies
   in the shell $\Lambda_{l+dl}<|\varepsilon_p(\bm k)|< \Lambda_l$,
  while the crossed line represents the electron having high energies 
  determined by the momentum conservation.
The diagrams where the crossed line  and slashed line are interchanged 
  are also taken into account.
}
\label{fig:vertex1}
\end{figure}
%======================================================================

First we focus on the one-loop contributions due to the 
   second-order vertex corrections.  
Possible Peierls and Cooper bubble contributions,
  due to the normal and umklapp scattering,
  are shown in Fig.\ \ref{fig:vertex1}.
We integrate out the electron degrees of freedom which have
  energy  in the shell $\Lambda_{l+dl}<|\varepsilon_p(\bm k)|< \Lambda_l$.
The respective   Peierls and Cooper bubbles have 
  the \textit{transverse}-momentum (i.e., patch-index) dependence 
  of the external variables, 
  as discussed in the literature.
  \cite{Duprat2001,Bourbonnais2004,Fuseya2005,Dupuis2005}
This effect is crucial to induce the transverse-momentum dependence of the 
  coupling constants.
There remain ambiguities in the selection of the 
  \textit{longitudinal} momenta for the external variable, 
  since, in general, all the momenta of vertex cannot be set on the 
  Fermi surface  if the Fermi surface is warped. \cite{Doucot2003}
In this paper,  we set three of four external momenta being on the
  Fermi surface and the longitudinal momentum conservation 
  for each vertex (even for the internal momenta) is also considered.
The choice of the external longitudinal momenta,
   in addition to the transverse momenta,
 affects on the internal 
   momenta and also on the RG equations.
To keep the symmetries discussed in the preceding section,
  we also take into account the different choice of  
  three of four longitudinal momenta on the Fermi surface.
The explicit form of the  Peierls bubble is given by
  $-(T/V)\sum_{\bm k}^{\mathrm{o.s.}} \sum_n 
  \mathcal{G}_{0+}(\bm k,i\omega_n)
  \mathcal{G}_{0-}(\bm k-\bm q, i\omega_n) $
   where 
   $\mathcal{G}_{0p}(\bm k,i\omega_n)
    =[i\omega_n - \varepsilon_p(\bm k)]^{-1}$
 is the Green's function
  for the noninteracting case.
By taking  summation of the Matsubara frequency  and 
  by performing the outer-shell integral over constant energy,  
  this Peierls bubble contribution  is given by 
  $(2\pi v N_\perp)^{-1} \sum_{k} I_{(q,k,k_1,k_2)} dl$, where
  the cutoff function $I_{(q,k,k_1,k_2)}$ is given  in the $T\to 0$
   limit by
%====================================================================
\begin{eqnarray}
I_{(q,k,k_1,k_2)}
=
\frac{\Lambda}{2}\sum_{p=\pm}  \sum_{i=1,2}
\frac{\Theta\biglb(\Lambda+pA_{q,k,k_i}(l)\bigrb)} 
      {2\Lambda+pA_{q,k,k_i}(l)} .
\label{eq:I}
\end{eqnarray}
%====================================================================
The quantity $A_{q,k,k'}(l)$ 
 being the functions of $t_\perp(l)$ is given by
%====================================================================
\begin{eqnarray}
A_{q,k,k'}(l)&\equiv&
2t_\perp(l) [\cos k +\cos (k-q)]
\nonumber \\ && {}
-2t_\perp(l) [\cos k' +\cos (k'-q)].
\label{eq:A}
\end{eqnarray}
%====================================================================
The second term in the rhs of $A_{q,k,k'}(l)$ appears due to 
the longitudinal momentum conservation. 
In the conventional approach, this  term has been 
  neglected, \cite{Duprat2001,Bourbonnais2004,Fuseya2005,Dupuis2005}
  however is crucial to 
  reproduce the known RG equations in the two-leg ladder system
  (Sec.\ \ref{sec:ladder}).
We note $I_{(q,k,k_1,k_2)}=1$ in the 1D limit ($t_\perp\to 0$). 
The Cooper bubble contribution is also calculated in 
  a similar way and  can be expressed,
  after some algebra,  as $-I_{(\pi-q+k_1+k_2,k',k_1,k_2)}$
 where we have used the particle-hole symmetry.

%====================================================================
\begin{figure}[t]
\includegraphics[width=7.5cm]{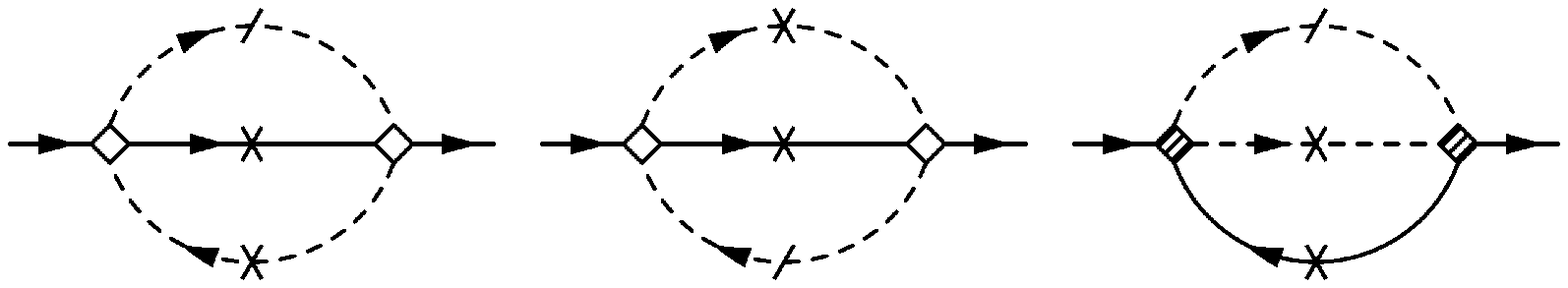}
\caption{
The logarithmic-singular 
  second-order diagrams for the Green's function, contributing 
  the self energy.
The notations are the same as in Fig.\ \ref{fig:vertex1}.
Other types of second-order diagrams 
  are not logarithmic singular even in the 1D limit  and can be neglected.
}
\label{fig:selfenergy}
\end{figure}
%======================================================================

\subsection{Two-loop self-energy corrections}

To go beyond the one-loop RG theory, we have to take into
  account two-loop self-energy corrections based on the
  second-order perturbation.
The Fermi surface deformation
  can be taken into account by considering these corrections
  and has been discussed intensively
  by Dusuel and Dou\c{c}ot, \cite{Doucot2003}
  based on the zero-temperature formalism.
Here we perform the finite-temperature formalism 
  and take the $T\to 0$ limit at the final stage of the calculation.
The second-order self-energy diagrams are shown in Fig.\ \ref{fig:selfenergy}.
In the second-order perturbation,
  there are two types of corrections to the
  single-particle Green's function $\mathcal{G}$:
One is the corrections to the wave-function renormalization factor 
  while the other 
  contributes to the renormalization of the velocity and the 
  interchain hopping. 
In the present RG scheme, the renormalization factor can have
  a transverse momentum dependence.
So we assume that the Green's function takes a form 
%====================================================================
\begin{eqnarray}
  \mathcal{G}_{p}(\bm k,i\omega_n) 
=
\frac{z_{k_\perp}^p}{i\omega_n - v(p k_\parallel-k_F) 
 + 2 t_{\perp}^{\mathrm{eff}}  \cos k_\perp}
.
\label{eq:Green_general}
\end{eqnarray}
%====================================================================
where $z^R_{k_\perp}=z^L_{-k_\perp} (\equiv z_{k_{\perp}})$.
The explicit calculation of outer shell integration of the 
  diagrams in Fig.\ \ref{fig:selfenergy} yields
\begin{widetext}
%====================================================================
\begin{eqnarray}
\mathcal{G}_{R}^{-1}(\bm k,i\omega_n) 
&=&
\mathcal{G}_{0R}^{-1}(\bm k,i\omega_n)
-
\frac{dl}{2N_\perp^2} \sum_{q , k'}  G_{\Sigma(q,k,k')}^2
\left[ 
J_{0(q,k,k')}
-
J_{1(q,k,k')} \mathcal{G}_{0R}^{-1}(\bm k,i\omega_n) 
\right],
\label{eq:selfenergy}
\end{eqnarray}
%====================================================================
for the right-moving electrons.
The second-order coupling constants contributing the
  self-energy corrections are put into a form:
%====================================================================
\begin{eqnarray}
G_{\Sigma(q,k,k')}^2 
&\equiv &
G_{1\perp(q,k,k')}^2
+ G_{2\perp(q,k,k')}^2
+ G_{\parallel(q,k,k')}^2
+ \frac{1}{2} G_{3\perp(q,k,\pi-k')}^2
+ \frac{1}{2} G_{3\perp(\pi-q+k+k',k,\pi-k')}^2
\nonumber \\ && {}
+ \frac{1}{2} G_{3\parallel(q,k,\pi-k')}^2
- G_{3\parallel(q,k,\pi-k')}G_{3\parallel(\pi-q+k+k',k,\pi-k')}
+ \frac{1}{2} G_{3\parallel(\pi-q+k+k',k,\pi-k')}^2
 .
\label{eq:self-energy}
\end{eqnarray}
%====================================================================
\end{widetext}
We note that the umklapp scattering with the same spins $G_{3\parallel}$
  also has finite contributions  which are absent in the 1D limit.
The quantities $J_{0(q,k,k')}$ and $J_{1(q,k,k')}$ 
  denote the cutoff functions due to the warped Fermi surface,
  which are also determined by the quantity 
   $A_{q,k,k'}(l)$ [Eq.\ (\ref{eq:A})].
These cutoff functions $J_0$ and $J_1$ take different forms
  depending on the relation between $A_{q,k,k'}(l)$ and $\Lambda$:
For $|A_{q,k,k'}(l)|<\Lambda$, these are given by
%====================================================================
\begin{subequations}
\begin{eqnarray}
J_{0(q,k,k')}
&=&
2\Lambda
\ln\left[
   \frac{4\Lambda+A_{q,k,k'}(l)}
        {4\Lambda-A_{q,k,k'}(l)}
\right],
\label{eq:J0}
\\
J_{1(q,k,k')}
&=&
 \frac{16\Lambda^2}{16\Lambda^2-A^2_{q,k,k'}(l)}.
\end{eqnarray}
\end{subequations}
%====================================================================
For $|A_{q,k,k'}(l)|>\Lambda$, 
%====================================================================
\begin{subequations}
\begin{eqnarray}
J_{0(q,k,k')}
&=&
2\Lambda
\ln\left[
   \frac{4\Lambda+|A_{q,k,k'}(l)|}{2\Lambda+|A_{q,k,k'}(l)|}
\right]  \, \mathrm{sgn} \biglb(A_{q,k,k'}(l) \bigrb),
\nonumber \\
\\
J_{1(q,k,k')}
&=&
 \frac{2\Lambda}{4\Lambda+|A_{q,k,k'}(l)|}
     + \frac{2\Lambda}{2\Lambda+|A_{q,k,k'}(l)|}.
\end{eqnarray}
\end{subequations}
%====================================================================
There remain subtleties in the integral region of outer shell,
   \cite{Bourbonnais2003} here we adopt the simplest shell integral
   following Ref.\ \onlinecite{Bourbonnais1991}.
The scaling deviation terms \cite{Bourbonnais2003,Bourbonnais1991}
 have been neglected.

The self-energy corrections proportional to
  $\mathcal{G}^{-1}_{0R}(\bm k , i\omega_n)$
  in Eq.\ (\ref{eq:selfenergy})  contribute to 
  the wave-function renormalization factor $z_{k_\perp}^p$.
The explicit RG equation  of the wave-function renormalization factor
  is given by 
%====================================================================
\begin{eqnarray}
\frac{d}{dl} \ln z_{k}
 &=&
- \frac{1}{2N_\perp^2} \sum_{q , k'}
 G_{\Sigma(q,k,k')}^{2} \,
J_{1(q,k,k')} 
.
\end{eqnarray}
%====================================================================
The self-energy corrections
   proportional to $J_0$  in Eq.\ (\ref{eq:selfenergy})  
 contribute to the renormalization of the velocity and the 
   Fermi surface deformation.
To simplify  discussions
  in the present analysis, we neglect the velocity renormalization, 
  since this effect would only yield quantitative changes.
The Fermi surface deformation can be extracted from these
  second-order corrections.
Since the Fermi surface is given by Eq.\ (\ref{eq:kf}),
  the Fermi surface deformation thus corresponds to 
  the renormalization of the interchain hopping.
By noting that the self-energy  contributions to 
  the interchain hopping should have
   transverse momentum  dependence $\cos k$, 
  the RG equation of the renormalization for the interchain hopping
 is given by
%====================================================================
\begin{eqnarray}
\frac{d}{dl}
t_\perp(l) &=&
t_\perp(l) 
\nonumber \\ &&{} \hspace*{-1cm}
-
\frac{1}{4N_\perp^3} \sum_{q ,k, k'}
 G_{\Sigma(q,k,k')}^2 
\, J_{0(q,k,k')} \cos k.
\label{eq:RG_tperp}
\end{eqnarray}
%====================================================================
The renormalization to higher-order interchain
  hopping has been neglected.

\subsection{Two-loop RG equations}

%====================================================================
\begin{figure}[t]
\includegraphics[width=8cm]{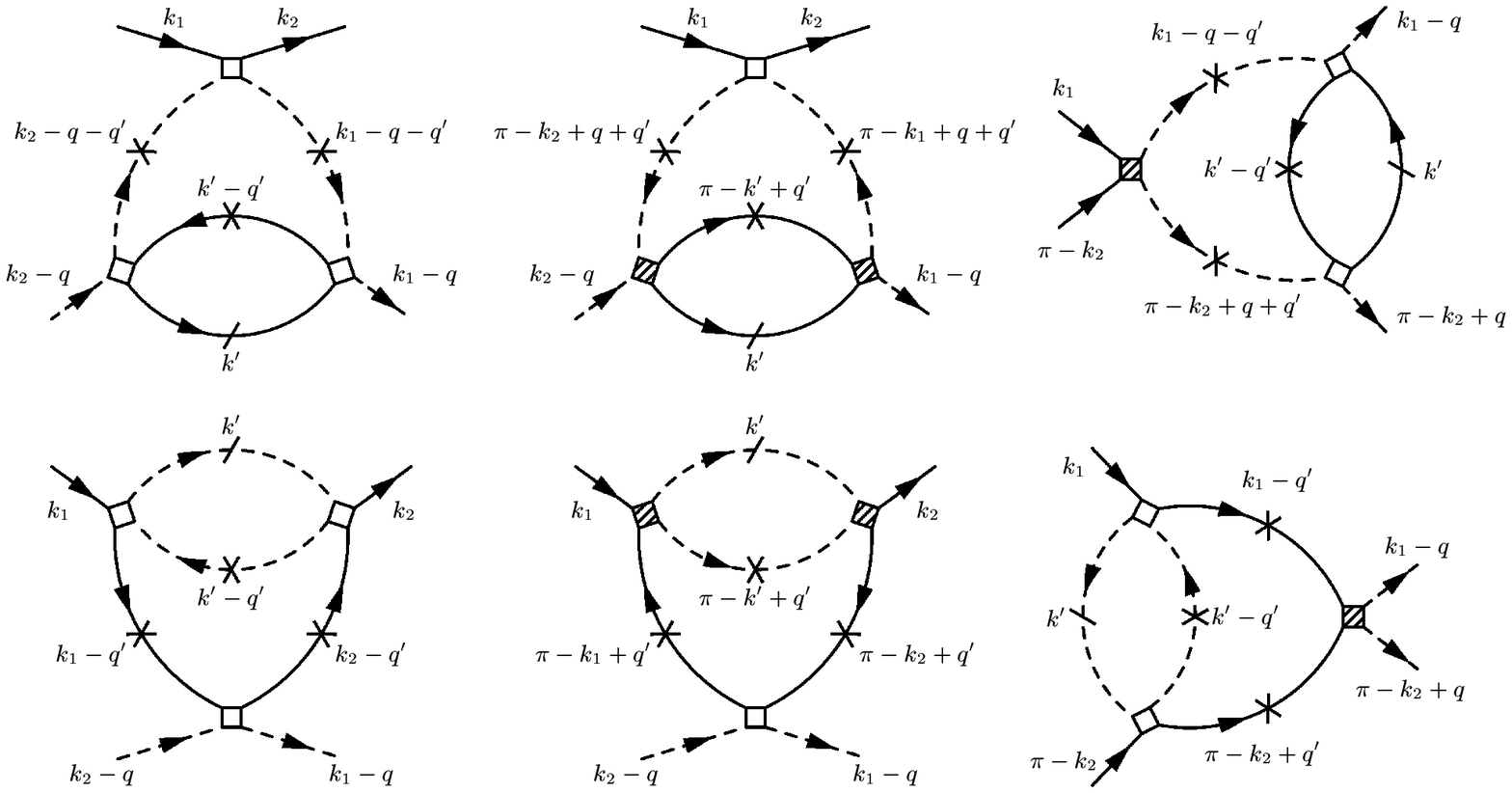}
\caption{
The third-order diagrams contributing to the vertex corrections,
 which have an order $O(G^3 dl)$ in the 1D limit.
The notations are the same as in Fig.\ \ref{fig:vertex1}.
Other types of third-order diagrams have an order $O(G^3 dl^2)$
  and can be neglected.
}
\label{fig:vertex2}
\end{figure}
%======================================================================

In order to complete the two-loop RG theory, one has to take into
  account the next-to-leading logarithmic contributions to the
  vertex part.
The two-loop vertex corrections can be calculated in a similar way 
  to that for the self-energy correction.
The third-order diagrams with the next-to-leading logarithmic contributions
  are shown in Fig.\ \ref{fig:vertex2} and yield the renormalization 
  of the vertex as 
  $G_{i(q,k_1,k_2)} \to z_{i(q,k_1,k_2)} \, G_{i(q,k_1,k_2)}$,
  where $i=1$$\perp$, $2$$\perp$, $\parallel$, $3$$\perp$, and $3$$\parallel$.
Other types of diagrams are of the order $O(G^3 dl^2)$ which are already
  taken into account in the one-loop level.
As is well-known in the 1D case, the RG is formulated 
  by deriving the scaling equations for 
  the ``renormalized'' coupling constants
   $G_{i}(l) \equiv  G_{i} z_{i}(l) z^2(l)$
  \cite{Bourbonnais2003,Solyom,Bourbonnais1991}
  where $z(l)$ is the wave-function renormalization factor.
In the present Q1D RG, 
  by keeping in mind that the vertex has a transverse-momentum dependence,
the renormalized coupling constants are
   defined as
%====================================================================
\begin{eqnarray}
G_{i(q,k_1,k_2)}(l)
&\equiv&
 G_{i(q,k_1,k_2)} \,
   z_{i(q,k_1,k_2)}(l) 
\nonumber \\ && {} \times
 \sqrt{
  z_{k_1}^R(l) \, z_{k_2}^R(l) \, 
  z_{k_1-q}^L(l) \, z_{k_2-q}^L(l) \, 
  },
\nonumber \\
\end{eqnarray}
%====================================================================   
for the normal scatterings $(i=1\!\perp,2\!\perp,\parallel)$, and 
%====================================================================
\begin{eqnarray}
G_{i(q,k_1,k_2)}(l)
&\equiv&
 G_{i(q,k_1,k_2)} \,
   z_{i(q,k_1,k_2)}(l) 
\nonumber \\ && {} \times
 \sqrt{
  z_{k_1}^R(l) \, z_{k_2}^R(l) \, 
  z_{k_1-q}^L(l) \, z_{k_2+q}^L(l) \, 
  },
\nonumber \\
\end{eqnarray}
%====================================================================   
for the umklapp scatterings $(i=3\!\perp,3\!\parallel)$.
The wave-function renormalization factor $z_{k_\perp}^p$ comes from the
  rescaling of the electron field operator.
Even in the two-loop vertex corrections, the cutoff function 
  due to the warping of the Fermi surface appears, which is given by
%====================================================================
\begin{eqnarray}
J_{2(q+k'';k_1,k_2;k',k'')}
&=&
\frac{1}{2}J_{1(q+k''-k_1,k',k'')}
\nonumber \\ && {}
+\frac{1}{2}J_{1(q+k''-k_2,k',k'')}.
\end{eqnarray}
%====================================================================
The cutoff function $I$, $J_{0}$,  $J_{1}$, and $J_{2}$
  are not universal and would take different
  forms depending on the RG formulation.
The function $I$ [Eq.\ (\ref{eq:I})]
  is not continuous as a function of $A_{q,k,k_i}(l)$ which 
  would be due to the sharp cutoff of the bandwidth.
This unphysical discontinuity of $I$ affects the 
  results of the numerical integration of the RG equations. 
In order to avoid this unphysical effect,
  we replace $I$ by a smooth function which 
  reproduce the limiting behavior of Eq.\ (\ref{eq:I})
  for small and large $A_{q,k,k_i}(l)$.

From the straightforward calculation of the diagrams in Fig.\
  \ref{fig:vertex2}, we obtain the two-loop
  RG equations for $G_{1\perp(q,k_1,k_2)}$,
  $G_{2\perp(q,k_1,k_2)}$,
  $G_{\parallel(q,k_1,k_2)}$,
  $G_{3\perp(q,k_1,k_2)}$, and
  $G_{3\parallel(q,k_1,k_2)}$.
We note that, if we set $N_\perp=2$
  and if we neglect the umklapp scattering
  $G_{3\perp(q,k_1,k_2)}$ and
  $G_{3\parallel(q,k_1,k_2)}$, our RG equations
   reproduce the two-loop RG equations
  obtained by Fabrizio \cite{Fabrizio1993}
  in the two-leg ladder system at away from half filling.
By using Eq.\ (\ref{eq:notation}),  we rewrite 
  the RG equation in terms of 
  $G_\rho$, $G_\sigma$, $G_c$, $G_s$, and $G_{cs}$.
For the system with the spin-rotational SU(2) symmetry,
  the coupling constants satisfy the relations given by
   Eq.\ (\ref{eq:spinSU(2)_gology}).
The full RG equations in this case is given in the Appendix
  \ref{sec:appendix_RG}.
For the particle-hole symmetric Hubbard model,
   the coupling constants also satisfy the relations 
   (\ref{eq:ph}) and (\ref{eq:chargeSU(2)_gology}).
By using all these relations,
  the RG equations with the SU(2) $\times$ SU(2) symmetry
  are extremely simplified.
The complete two-loop RG equations for the coupling constants
  $G_{\rho(q,k_1,k_2)}$ and  $G_{\sigma(q,k_1,k_2)}$ are given by
\begin{widetext}
%====================================================================
\begin{eqnarray}
\frac{d}{dl}
G_{\nu(q,k_1,k_2)}
&=&
\frac{1}{2N_\perp} \sum_{k'}
\left[
    \alpha_{\nu(q;k_1,k_2;k')}  I_{(q,k',k_1,k_2)}
  - \beta_{\nu(q;k_1,k_2;k')}  I_{(\pi-q+k_1+k_2 ,k',k_1,k_2)}
\right]
\nonumber \\ && {} %---------
- \frac{1}{4N_\perp^2}  \,  G_{\nu(q,k_1,k_2)} 
\sum_{q' ,k'}  
\Bigl[
  G_{\Sigma (q',k_1,k')}^2 \, J_{1(q',k_1,k')}
+ G_{\Sigma (q',k_2,k')}^2 \, J_{1(q',k_2,k')}
\Bigr]
\nonumber \\  && {}
- \frac{1}{4N_\perp^2}  \,  G_{\nu(q,k_1,k_2)} 
\sum_{q' ,k'}  
\Bigl[
  G_{\Sigma (q',-k_1+q,k')}^2 \, J_{1(q',-k_1+q,k')}
+ G_{\Sigma (q',-k_2+q,k')}^2 \, J_{1(q',-k_2+q,k')}
\Bigr]
\nonumber \\ && {} %---------
+ \frac{1}{4N_\perp^2} \sum_{q',k'}
\Bigl\{
   \bigl[
     G_{\nu(q+q',k_1,k_2)} 
     - \Theta_\nu G_{\nu(\pi-q-q'+k_1+k_2,k_1,k_2)}
   \bigr]
   \gamma_{\nu(q-k_1+k',q-k_2+k';k',k';q')}
\nonumber \\ && {} \qquad\qquad
- \frac{1}{2}
  G_{\nu(\pi-q-q'+k_1+k_2,k_1,k_2)}
  \delta_{\nu(q-k_1+k',q-k_2+k';k',k';q')}
\Bigr\}
  J_{2(q+k';k_1,k_2;k',k'-q')}
\nonumber \\ && {}
+ \frac{1}{4N_\perp^2} \sum_{q',k'}
\Bigl\{
   \bigl[
     G_{\nu(q-q',k_1-q',k_2-q')} 
     - \Theta_\nu G_{\nu(\pi-q-q'+k_1+k_2,k_1-q',k_2-q')}
   \bigr]
   \gamma_{\nu(k_1-k',k_2-k';k_1,k_2;q')}
\nonumber \\ && {} \qquad\qquad
- \frac{1}{2}
  G_{\nu(\pi-q-q'+k_1+k_2,k_1-q',k_2-q')}
  \delta_{\nu(k_1-k',k_2-k';k_1,k_2;q')}
\Bigr\}
  J_{2(k';k_1,k_2;k',k'-q')}
,
\label{eq:RG_G}
\end{eqnarray}
%====================================================================
where $\nu=\rho,\sigma$ and the sign function $\Theta_\nu$ is 
$\Theta_\rho=+1$ and $\Theta_\sigma=-1$.
The index of the scaling parameter $l$ in the coupling constants 
  $G_{\nu(q,k_1,k_2)}$   is suppressed.
The coupling constants for the self-energy corrections,
Eq.\ (\ref{eq:self-energy}),
 can be rewritten in terms of $G_\rho$ and $G_\sigma$, as
%====================================================================
\begin{equation}
 G_{\Sigma(q,k,k')}^2 
=
 G_{\rho(q,k,k')}^2  
+\frac{1}{2} G_{\rho(q,k,k')}  G_{\rho(\pi-q+k+k',k,k')}
+ 3 G_{\sigma(q,k,k')}^2 
- \frac{3}{2}  G_{\sigma(q,k,k')} G_{\sigma(\pi-q+k+k',k,k')}.
\label{eq:Gself}
\end{equation}
%====================================================================
The quantities $\alpha_\nu$, $\beta_\nu$, $\gamma_\nu$, and
 $\delta_\mu$ ($\nu=\rho,\sigma$)
 are  defined as follows.
The quantities $\alpha_\nu$ represent the one-loop 
  Peierls bubble contributions given by
%====================================================================
\begin{eqnarray*}
\alpha_{\rho (q;k_1,k_2;k')}
&\equiv&
  2G_{\rho(q,k_1,k')} \,  G_{\rho(q,k',k_2)}
+  G_{\rho(q,k_1,k')} \,  G_{\rho(\pi-q+k_2+k',k',k_2)}
+  G_{\rho(\pi-q+k_1+k',k_1,k')} \, G_{\rho(q,k',k_2)}   
\nonumber \\ && {}
+ 6G_{\sigma(q,k_1,k')} \, G_{\sigma(q,k',k_2)}
 -3G_{\sigma(q,k_1,k')} \, G_{\sigma(\pi-q+k_2+k',k',k_2)}
 -3G_{\sigma(\pi-q+k_1+k',k_1,k')} \, G_{\sigma(q,k',k_2)} 
\nonumber \\ && {}
+ G_{\rho(\pi-q+k_1+k',k_1,k')} \,
  G_{\rho(\pi-q+k_2+k',k',k_2)}
+ 3 G_{\sigma(\pi-q+k_1+k',k_1,k')} \,
    G_{\sigma(\pi-q+k_2+k',k',k_2)},
\\
\alpha_{\sigma (q;k_1,k_2;k')}
&\equiv&
 2G_{\rho(q,k_1,k')}  \,  G_{\sigma(q,k',k_2)}
+2G_{\sigma(q,k_1,k')} \, G_{\rho(q,k',k_2)}
-4G_{\sigma(q,k_1,k')} \, G_{\sigma(q,k',k_2)} 
\nonumber \\ && {}
 - G_{\rho(q,k_1,k')}   \, G_{\sigma(\pi-q+k_2+k',k',k_2)}
 - G_{\sigma(\pi-q+k_1+k',k_1,k')} \, G_{\rho(q,k',k_2)} 
 + 2G_{\sigma(q,k_1,k')} \, G_{\sigma(\pi-q+k_2+k',k',k_2)}
\nonumber \\ && {}
 + G_{\rho(\pi-q+k_1+k',k_1,k')} \, G_{\sigma(q,k',k_2)} 
 + G_{\sigma(q,k_1,k')} \, G_{\rho(\pi-q+k_2+k',k',k_2)} 
 + 2 G_{\sigma(\pi-q+k_1+k',k_1,k')} \, G_{\sigma(q,k',k_2)} 
\nonumber \\ && {}
- G_{\rho(\pi-q+k_1+k',k_1,k')} \, G_{\sigma(\pi-q+k_2+k',k',k_2)}
- G_{\sigma(\pi-q+k_1+k',k_1,k')} \, G_{\rho(\pi-q+k_2+k',k',k_2)} 
\nonumber \\ && {}
- 2 G_{\sigma(\pi-q+k_1+k',k_1,k')} \, G_{\sigma(\pi-q+k_2+k',k',k_2)}.
\end{eqnarray*}
%====================================================================
The quantities $\beta_\nu$ represent the one-loop 
  Cooper bubble contributions:
%====================================================================
\begin{eqnarray*}
\beta_{\rho(q;k_1,k_2;k')}
&\equiv&
    G_{\rho(q-k_2+k',k_1,k')} G_{\rho(q-k_1+k', k',k_2)}
+ 3 G_{\sigma(q-k_2+k',k_1,k')} G_{\sigma(q-k_1+k', k',k_2)} ,
\\
\beta_{\sigma(q;k_1,k_2;k')}
&\equiv&
   G_{\rho(q-k_2+k',k_1,k')} G_{\sigma(q-k_1+k', k',k_2)}
 + G_{\sigma(q-k_2+k',k_1,k')} G_{\rho(q-k_1+k',k',k_2)} 
\nonumber \\  && {}
 +2G_{\sigma(q-k_2+k',k_1,k')} G_{\sigma(q-k_1+k', k',k_2)}.
\end{eqnarray*}
%====================================================================
Finally the quantities $\gamma_\nu$ and $\delta_\nu$
  represent the two-loop vertex contributions:
%====================================================================
\begin{eqnarray*}
\gamma_{\rho (q_1,q_2;k_1,k_2;q')}
&\equiv&
   G_{\rho  (q_1,k_1,k_1-q')} \, G_{\rho  (q_2,k_2,k_2-q')}
+ 3G_{\sigma(q_1,k_1,k_1-q')} \, G_{\sigma(q_2,k_2,k_2-q')}  ,
\\
\gamma_{\sigma(q_1,q_2;k_1,k_2;q')}
&\equiv&
   G_{\rho  (q_1,k_1,k_1-q')} \,  G_{\rho  (q_2,k_2,k_2-q')}
-  G_{\sigma(q_1,k_1,k_1-q')} \,  G_{\sigma(q_2,k_2,k_2-q')} ,
\\
\delta_{\rho (q_1,q_2;k_1,k_2;q')}
&\equiv&
   G_{\rho  (q_1,k_1,k_1-q')} \, G_{\rho  (\pi-q_2-q'+2k_2,k_2,k_2-q')}
+  G_{\rho  (\pi-q_1-q'+2k_1,k_1,k_1-q')} \, G_{\rho  (q_2,k_2,k_2-q')}
\nonumber  \\ && {}
-3 G_{\sigma(q_1,k_1,k_1-q')} \, G_{\sigma(\pi-q_2-q'+2k_2,k_2,k_2-q')}  
-3 G_{\sigma(\pi-q_1-q'+2k_1,k_1,k_1-q')} \, G_{\sigma(q_2,k_2,k_2-q')}  ,
\\
\delta_{\sigma(q_1,q_2;k_1,k_2;q')}
&\equiv&
 - G_{\rho  (q_1,k_1,k_1-q')} \, G_{\rho  (\pi-q_2-q'+2k_2,k_2,k_2-q')}
 - G_{\rho  (\pi-q_1-q'+2k_1,k_1,k_1-q')} \, G_{\rho  (q_2,k_2,k_2-q')}
\nonumber  \\ && {}
 - G_{\sigma(q_1,k_1,k_1-q')} \, G_{\sigma(\pi-q_2-q'+2k_2,k_2,k_2-q')}  
 - G_{\sigma(\pi-q_1-q'+2k_1,k_1,k_1-q')} \, G_{\sigma(q_2,k_2,k_2-q')}  .
\end{eqnarray*}
%====================================================================
\end{widetext}
We have only kept the marginal scattering processes.
In the purely 1D case, it is known that the $G_{cs}$ term has
  irrelevant canonical dimension. \cite{Tsuchiizu_Furusaki}
In the present case, some of the $G_{cs(q,k_1,k_2)}$ couplings 
  have a marginal canonical dimension, however,  
  the RG equation for the $G_{cs(q,k_1,k_2)}$ 
  does not appear explicitly since
  the correction due to this term always appears
  in a form $(G_{cs(q,k_1,k_2)}-G_{cs(\pi-q+k_1+k_2,k_1,k_2)})$,
  which shows the same $l$-dependence of 
 $(-G_{\sigma(q,k_1,k_2)}+G_{\sigma(\pi-q+k_1+k_2,k_1,k_2)})$,
 as seen from
   Eqs.\ (\ref{eq:spinSU(2)gs_gsigma}) and (\ref{eq:chargeSU(2)gs_gcs}).
For $N_\perp=8$,  e.g.,
the number of independent coupling constants reduces to 300 
  instead of 1440 for  without assuming the symmetries.
If the transverse momentum dependences of the coupling constants
  are neglected, the 1D RG equations [Eq.\ (\ref{eq:RG1d})]
 are reproduced.

From the numerical integration of the RG equations,  
  we can estimate characteristic energy scales.
Here we focus on the renormalized interchain hopping and
 the charge/spin excitation gaps.
The effective renormalized interchain hopping is given by
%====================================================================
\begin{eqnarray}
t_\perp^{\mathrm{eff}}
\equiv
\Lambda \exp (-l_{\perp}),
\end{eqnarray}
%====================================================================
where the quantity $l_\perp$ is determined from
 $t_\perp(l_\perp)=\Lambda$.
In the noninteracting limit, the interchain hopping scales
  as $t_\perp(l)=t_\perp e^{l}$, then $l_\perp= \ln(\Lambda/t_\perp)$ 
  and the effective interchain hopping trivially reduces to 
  the bare interchain hopping $t_\perp^{\mathrm{eff}}=t_\perp$.
This quantity characterizes the dimensional-crossover energy scale,
  below which the system cannot be regarded 
   as a one-dimensional system any more.
In addition, the Fermi surface deformation can be determined by 
  this quantity. 
By noting the relation (\ref{eq:kf}), the deformed Fermi surface is given by
%====================================================================
\begin{equation}
k_F^{\mathrm{eff}}(k_\perp)  =  
k_F +2 \frac{t_\perp^{\mathrm{eff}}}{v} \cos k_\perp.
\end{equation}
%====================================================================
It is known that the Fermi-surface 
  deformation comes only from the renormalization in the high-energy
  regime, since the
  coupling constants which appear in 
  the rhs of Eq.\ (\ref{eq:RG_tperp}) are the irrelevant couplings.
  \cite{Doucot2003}

\begin{figure}[t]
\includegraphics[width=6.5cm]{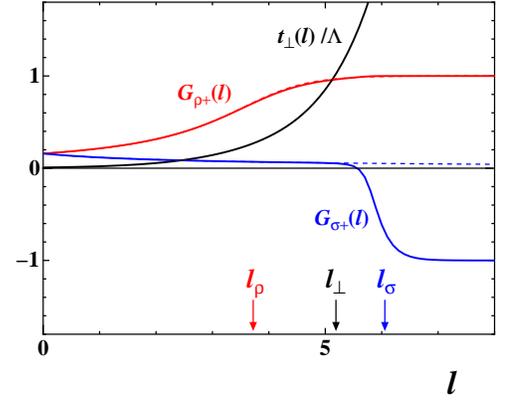}
\caption{
(Color online)
The scaling flows of the coupling constants $G_{\rho+}(l)$ and  
  $G_{\sigma+}(l)$ and the interchain hopping $t_\perp(l)/\Lambda$
 for $N_\perp=8$ 
  with fixed $U/t_\parallel = 2$ and $t_\perp/t_\parallel=0.05$.
The case for $t_\perp=0$ is shown by the dotted lines.
}
\label{fig:flow8}
\end{figure}

\begin{figure}[t]
\includegraphics[width=6.5cm]{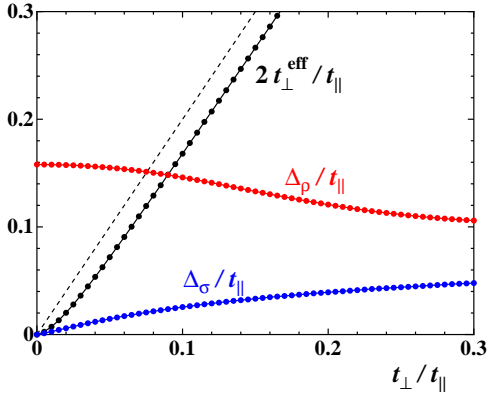}
\caption{
(Color online)
The charge gap $\Delta_\rho$, the spin gap $\Delta_\sigma$,
  and the characteristic energy scale  $t_\perp^{\mathrm{eff}}$,
   as a function of $t_\perp/t_\parallel$ for $N_\perp=8$ and 
  $U/t_\parallel = 2$.
The dashed line represents the magnitude of the bare interchain hopping.
}
\label{fig:Gap8}
\end{figure}

In the present RG scheme,
 the information of the charge gap $\Delta_\rho$  and the spin 
  gap $\Delta_\sigma$
can be extracted
  by focusing on the combination of the coupling constants: 
%====================================================================
\begin{eqnarray}
G_{\nu +}
&\equiv& 
\frac{1}{N_\perp^2} \sum_{k,k'}
 G_{\nu(k-k',k,k)} ,
\label{eq:gnu+}
\end{eqnarray}
%====================================================================
where $\nu=\rho, \sigma$. 
This interpretation can be justified by noting that
  the uniform charge/spin susceptibility 
  is determined by these quantities 
  \cite{Fuseya2006}
  and by the field-theoretical approach
  for the two-leg ladder ($N_\perp=2$) case
  as will be shown in Sec.\ \ref{sec:ladder}.
A typical scaling flow is shown in Fig.\ \ref{fig:flow8},
  where we have set $N_\perp=8$.
As a reference, the scaling flow for the 1D case is also shown.
The charge coupling $G_{\rho+}$ shows similar behavior 
  to that in the 1D case, while the spin coupling $G_{\sigma+}$ 
  becomes relevant and have a finite fixed point value $G_{\sigma-}^*=-1$.
We note that 
  the magnitude of several coupling constants becomes large and exceed the
  unity  under the scaling procedure
  for $l>l_\perp$.
By focusing on  this scaling behavior of  $G_{\nu+}(l)$, we can estimate
 the magnitude of the excitation gaps by
%====================================================================
\begin{eqnarray}
\Delta_\nu 
\equiv
\Lambda \exp (-l_{\nu})
\label{eq:gapestimation}
\end{eqnarray}
%====================================================================
where the quantity $l_\nu$ is determined from
 $|G_{\nu+}(l_\nu)|= c$ where $c$ is a numerical constant.
In the present numerical calculations,
  we will  set $c=0.7$ and $\Lambda=2v k_F$.
As seen in Eq.\ (\ref{eq:1DMottgap}), 
  these ambiguity simply affects on 
  the numerical factor and our choice 
  reproduce well  the exact results of the $\Delta_\rho$
  in the 1D case.\cite{Ovchinikov}
The interchain-hopping dependence of $\Delta_\rho$, $\Delta_\sigma$, 
  and $t_\perp^{\mathrm{eff}}$ is shown in Fig.\ \ref{fig:Gap8}.
The charge gap  
  is suppressed due to the  interchain hopping  
  but is always finite even when the interchain hopping 
  exceeds the magnitude of the charge gap.
In the present bipartite Q1D half-filled Hubbard model,
  we find that the  charge gap is always finite for  $U>0$.
This is contrast to the
   results obtained from the chain-DMFT 
  \cite{Giamarchi2001,Giamarchi2002,Giamarchi2004,Berthod}  
   where the metal-insulator (Mott) transition  has been 
  suggested for finite interchain hopping at $T=0$.
This difference would arise from the difference in the treatment  
  of the Fermi-surface nesting of the system.
In the present model, 
  the Fermi surface is always nested perfectly even for the
  finite interchain hopping where the nesting vector is $(\pi,\pi)$.
In our approach, we fully take into account this effect, 
 however in the chain-DMFT, 
  the warping of the Fermi surface is not taken into account.
We expect that 
  the Fermi-surface nesting would play crucial roles in 
  the Q1D Mott transition, since 
  the 1D Mott insulator itself is realized 
  even in the small $U$ region
  due to the commensurability effect, 
  which would be sensitive to the Fermi-surface nesting. 
By means of the present Q1D RG scheme,
  the effect of the nesting deviation will be reported elsewhere.
  \cite{Tsuchiizu_unpublished}

%====================================================================
%======================================================================

%====================================================================
%======================================================================

\section{Two-leg ladder model}\label{sec:ladder}

To indicate the validity of the two-loop RG equations obtained in the
  preceding section, we apply it to the two-leg Hubbard ladder model
   with a bipartite lattice.
This model has been investigated by the RG method combined with 
  the analytical field-theoretical method
  \cite{Nersesyan1993,Khveshchenko1994,Schulz1996,Balents1996,Lin1998}
  and by the numerical DMRG method, 
  \cite{Noack,Weihong}
 and it has been clarified that
  the spin-gapped insulating state  called the $D$-Mott phase 
  is realized.
Lin, Balents and Fisher obtained the highly symmetric SO(8) Gross-Neveu model 
  as an effective theory in the low-energy limit 
  by using the fixed-point behavior of the one-loop RG analysis.
  \cite{Lin1998}
They further discussed finite-energy spectrum based on this effective
  theory, however,  it is not clear that this high symmetry still 
  holds at finite-energy scale.
Actually, the RG method allows us to study the characteristic energy scales 
 in addition to the fixed point behavior,
  however,  the naive one-loop RG is not sufficient 
  to estimate the excitation gaps,  since the RG method breaks 
  down at the scale corresponding to the largest gap, as mentioned before. 
A promising method is to derive an effective theory by 
  tracing out the gapped modes
  based on the field-theoretical treatment.
However, in the present two-loop RG, 
  the excitation gaps in the respective modes 
  can be estimated without following the tracing-out procedure.
This is not so trivial if
  the respective modes are not independent.
In this section,
  in order to check the validity of the present method,
  we consider the two-leg ladder system, which is a minimal model
  of the spin-charge coupled systems,
  and confirm that this two-loop RG theory reproduces 
  results obtained by the DMRG method
  and further analyze the excitation properties in detail 
  by combining the field-theoretical bosonization and  
  fermionization method.

The model can be obtained from Eq.\ (\ref{eq:model}) 
  by simply setting  $N_\perp=2$.
The possible values of the transverse momentum are $k_\perp=0$ and $\pi$.
From the symmetry requirements 
  [Eqs.\ (\ref{eq:spinSU(2)_gology}), (\ref{eq:ph}), and 
   (\ref{eq:chargeSU(2)_gology})],
the number of independent coupling constants reduces to 8  
  instead of 30 for  without assuming the symmetries.
To respect these symmetries and to make the physical picture 
  transparent,
  we derive the effective low-energy theory  by
  applying the bosonization and refermionization.
  \cite{Tsuchiizu2002,Tsuchiizu2005,Shelton}

First we apply the conventional Abelian bosonization to the Hamiltonian. 
The field operators of the right and left-moving electrons
are written as
%============================================================
\begin{equation}
\psi_{p,s,\zeta}(x) =
 \frac{\eta_{s,\zeta}}{\sqrt{2\pi a} }
\exp \left( ipk_{F,\zeta} x 
 + i p\, \varphi _{p,s,\zeta} \right),
\label{eq:field} 
\end{equation}
%===========================================================
   where $p=+$/$-$ represents the right/left moving electron,
   $s$ represents the spin, $\zeta$ represents the band
  index: $\zeta=+(-)$ for $k_\perp=0(\pi)$, 
  and $k_{F,\pm}=(\pi/2 \pm 2 t_\perp/v)$ [see Eq.\ (\ref{eq:kf})].
The technical details can be found in Refs.\ 
  \onlinecite{Tsuchiizu2002} and \onlinecite{Tsuchiizu2005}.
The chiral bosons obey the commutation relations
   $[\varphi_{p,s,\zeta}(x),\varphi_{p,s',\zeta'}(x')]
    = ip\pi \, \mathrm{sgn}(x-x') \, 
    \delta_{s,s'}\,\delta_{\zeta,\zeta'}$
   and
   $[\varphi_{+,s,\zeta},\varphi_{-,s',\zeta'}]
    = i\pi \,\delta_{s,s'}\,\delta_{\zeta,\zeta'}$.
The Klein factors $\eta_{s,\zeta}$, which satisfy
   $\{\eta_{s,\zeta},\eta_{s',\zeta'}\}
   =2\delta_{s,s'}\delta_{\zeta,\zeta'}$,
   are introduced in order to retain the correct anticommutation
   relation of the field operators between the different spin and
   band index.
To express the electron fields in terms of the bosonic fields 
  representing physical modes, we define a new
  set of chiral bosonic fields
  $\phi_{\rho +}^p$, $\phi_{\rho -}^p$, $\phi_{\sigma +}^p$, and
  $\phi_{\sigma -}^p$, by
%============================================================
\begin{equation}
\varphi_{p,s,\zeta}
  \equiv
     \phi_{\rho +}^p
   + \zeta  \phi_{\rho -}^p
   + s \phi_{\sigma +}^p
   + s \zeta \phi_{\sigma -}^p
,
\end{equation}
%===========================================================
    where $s=\uparrow$$/$$\downarrow=+/-$.
The commutation relations for these bosonic fields are
$[\phi_{\nu r}^p(x), \phi_{\nu' r'}^p(x')] 
= ip \, (\pi/4) \, 
   \mathrm{sgn}(x-x') \, \delta_{\nu,\nu'} \delta_{r, r'}$
 and
$[\phi_{\nu r}^+(x), \phi_{\nu' r'}^-(x')] 
=i \, (\pi/4) \,  \delta_{\nu,\nu'} \delta_{r, r'}$.
From Eq.\ (\ref{eq:field}) the density operator is given by
%========================================================== 
\begin{equation}
   :\! \psi_{p,s,\zeta}^\dagger\, \psi_{p,s,\zeta}^{} \! :
   \, = \,
   \frac{1}{2\pi} \, \frac{d}{dx} \varphi_{p,s,\zeta}(x)
.
\label{eq:density}
\end{equation}
%================================================================
The convention of  the Klein factors 
  is the same as Ref.\ \onlinecite{Tsuchiizu2002}.
From this relation, one finds that 
  the boson  fields $\phi_{\rho\pm}$
  can be interpreted to denote the ``charge'' degrees of freedom,  
  while $\phi_{\sigma\pm}$ to denote the ``spin'' degrees of freedom.

To appreciate two SU(2) symmetries in the effective theory,
 we next fermionize the $\phi_{\sigma+}$, $\phi_{\sigma-}$,
  and $\phi_{\rho+}$ bosonic fields by introducing
  the Majorana fermions
   $\xi^n_p$ ($n=1,\cdots,6$ and $p=R/L=+/-$):
%===============================================================
\begin{subequations}
\begin{eqnarray}
\frac{1}{\sqrt{2}} \left(\xi_{p}^2+i\xi_{p}^1\right)
 &\equiv& \frac{\kappa_{\sigma+}}{\sqrt{2\pi a}}
  \, \exp\left( ip \,2\phi_{\sigma+}^p \right),
\\
\frac{1}{\sqrt{2}} \left(\xi_{p}^4+i\xi_{p}^3\right)
 &\equiv& \frac{\kappa_{\sigma-}}{\sqrt{2\pi a}}
  \, \exp\left(ip \,2\phi_{\sigma-}^p\right),
\\
\frac{1}{\sqrt{2}} \left(\xi_{p}^6+i\xi_{p}^5\right)
 &\equiv& \frac{\kappa_{\rho+}}{\sqrt{2\pi a}}
  \, \exp\left( ip \,2\phi_{\rho+}^p\right),
\end{eqnarray}
\end{subequations}
%===============================================================
   where  $\kappa_{\nu \pm}$ is the Klein factor,
   satisfying $\{\kappa_{\nu r} , \kappa_{\nu' r'}\}=
   \delta_{\nu,\nu'}\delta_{r,r'}$ and
   $\kappa^2_{\nu r}=1$. 
These Majorana fields satisfy the anticommutation relations:
$ \{\xi_p^n(x),\xi_{p'}^{n'}(x')\}
=\delta(x-x') \, \delta_{p,p'} \, \delta_{n,n'} $.
The Hamiltonian  can be refermionized
 in terms of the Majorana fermions.
Our new finding is that
  the two sets of  three Majorana fields form triplets,
  due to the constraint of two SU(2) symmetries.
So we define
%===============================================================
\begin{eqnarray}
\bm\xi_p \equiv (\xi_p^1,\xi_p^2,\xi_p^3),\quad
\bm\zeta_p \equiv  (\xi_p^4,\xi_p^5,\xi_p^6).
\end{eqnarray}
%===============================================================
The $g$-ology Hamiltonian 
  $\int dx \, \mathcal{H}_{\mathrm{ladder}} 
  = (H_0+H_{\mathrm{I}})|_{N_\perp=2}$
  [Eqs.\ (\ref{eq:kinetic}) and (\ref{eq:g-ology}) with $N_\perp=2$]
  can be reexpressed in a highly symmetric form as 
%===============================================================
\begin{widetext}
\begin{eqnarray}
\mathcal{H}_{\mathrm{ladder}}
&=& 
-i\frac{v}{2}
\left(
  \bm\xi_{R} \cdot \partial_x \bm\xi_{R}
- \bm\xi_{L} \cdot \partial_x \bm\xi_{L}
\right)
-i\frac{v}{2}
\left(
  \bm\zeta_{R} \cdot \partial_x \bm\zeta_{R}
- \bm\zeta_{L} \cdot \partial_x \bm\zeta_{L}
\right)
+
\frac{v}{\pi}
\left[
   \left(\partial_x \phi_{\rho-}^R \right)^2 
  +\left(\partial_x \phi_{\rho-}^L \right)^2
\right]
\nonumber \\ && {}
- \frac{g_{\sigma +}}{2} 
\left( \bm\xi_{R} \cdot \bm\xi_{L} \right)^2
+ \frac{g_{\rho +}}{2} 
\left( \bm\zeta_{R} \cdot \bm\zeta_{L} \right)^2
+
\frac{g_{\rho-}}{\pi^2} 
   \left(\partial_x \phi_{\rho-}^R \right)
   \left(\partial_x \phi_{\rho-}^L \right)
\nonumber \\ && {}
- g_{\sigma-}
\left( \bm\xi_{R} \cdot \bm\xi_{L} \right)
\left( \bm\zeta_{R} \cdot \bm\zeta_{L} \right)
- \frac{ig_{\sigma(\pi,0,\pi)}}{2\pi a}
\left( \bm\xi_{R} \cdot \bm\xi_{L} \right)
 \cos 2 \theta_{\rho-} 
- \frac{ig_{\rho(\pi,0,\pi)}}{2\pi a}
\left( \bm\zeta_{R} \cdot \bm\zeta_{L} \right)
 \cos 2 \theta_{\rho-} 
\nonumber \\ &&{} 
+ \frac{ig_{\sigma(0,0,\pi)}}{2\pi a}
 \left( \bm\xi_{R} \cdot \bm\xi_{L} \right) \,
   \cos(2\phi_{\rho-}+8t_\perp x/v)
- \frac{ig_{\rho(0,0,\pi)} }{2\pi a}
 \left( \bm\zeta_{R} \cdot \bm\zeta_{L} \right) \,
  \cos(2\phi_{\rho-}+8t_\perp x/v)
,
\label{eq:Hladder}
\end{eqnarray}
\end{widetext}
%===============================================================
where 
$g_{\rho\pm } =
 \frac{1}{2}( g_{\rho(0,0,0)} \pm g_{\rho(\pi,0,0)})$ and
$g_{\sigma\pm} =
 \frac{1}{2}( g_{\sigma(0,0,0)} \pm g_{\sigma(\pi,0,0)} )$.
We note that the coupling constants $g_{\rho+}$ and $g_{\sigma+}$ are
  the same as defined in Eq.\ (\ref{eq:gnu+}).
From Eq.\ (\ref{eq:Hladder}), one easily finds that 
  the 6 Majorana fermions are not independent 
  and are grouped into 
 two triplets   $\bm\xi$ and $\bm\zeta$.
In the derivation of the above effective theory,
  we do not use any fixed point values of the coupling constants
  but simply have used symmetry constraints.
This means that the structure of the theory maintains
  at finite energy scale. 
The physical meanings of the respective triplets becomes clear by 
  noting the following relations.
The total spin operator $\bm S$ [Eq.\ (\ref{eq:spinoperator})]
  can be expressed in terms of the Majorana fermions 
  in a local form as
  $\bm S = \int dx \, \bm J(x)$ with
%====================================================================
\begin{eqnarray}
J^x(x)
&=&
-i 
\, \left( \xi^2_R \xi^3_R + \xi^2_L \xi^3_L \right),
\nonumber 
\\
J^y(x)
&=&
-i 
\, \left( \xi^3_R \xi^1_R + \xi^3_L \xi^1_L \right),
\label{eq:spincurrent}
\\
J^z(x)
&=&
-i 
\, \left( \xi^1_R \xi^2_R + \xi^1_L \xi^2_L \right).
\nonumber 
\end{eqnarray}
%===============================================================
Similarly
  the total ``charge'' operator $\bm Q$ [Eq.\ (\ref{eq:chargeoperator})]
  can be expressed as  $\bm Q = \int dx \, \bm J'(x)$ with
%====================================================================
\begin{eqnarray}
J'^x(x)
&=&
-i
\, \left( \xi^6_R \xi^4_R + \xi^6_L \xi^4_L \right),
\nonumber \\
J'^y(x)
&=&
-i
\, \left( \xi^4_R \xi^5_R + \xi^4_L \xi^5_L \right),
\label{eq:chargecurrent}
\\
J'^z(x)
&=&
-i
\, \left( \xi^5_R \xi^6_R + \xi^5_L \xi^6_L \right),
\nonumber 
\end{eqnarray}
%===============================================================
up to the Klein factor.
Thus we find that the system has the ``charge-triplet'' excitations
  described by the $\bm\zeta_p=(\xi_p^4,\xi_p^5,\xi_p^6)$
  Majorana fermions.
The derivation of these relations is quite similar to that 
  for the spin chains. \cite{Shelton}
These current operators satisfy the SU(2) Kac-Moody algebra at level 
  $k=2$. \cite{Gogolin}

For the relevant interchain hopping, we also find 
  high symmetry in the  $\rho-$ mode.
In this case, the terms $g_{\rho(0,0,\pi)}$ and $g_{\sigma(0,0,\pi)}$ 
  in Eq.\ (\ref{eq:Hladder})
  can be neglected due to the presence of $8t_\perp x/v$ in the 
  cosine potential
   and then 
 the effective theory  becomes
   SO(3)$\times$SO(3)$\times$U(1) symmetric,
where the  SO(3)$\times$SO(3) is due to the formation of
  two Majorana triplets and
  the U(1) is due to the absence of the potential
  for the bosonic field $\phi_{\rho-}$.
This picture is only valid for large interchain hopping,
  since the U(1) symmetry is retained dynamically 
  while the  SO(3)$\times$SO(3) has a microscopic origin.

\begin{figure}[t]
\includegraphics[width=7cm]{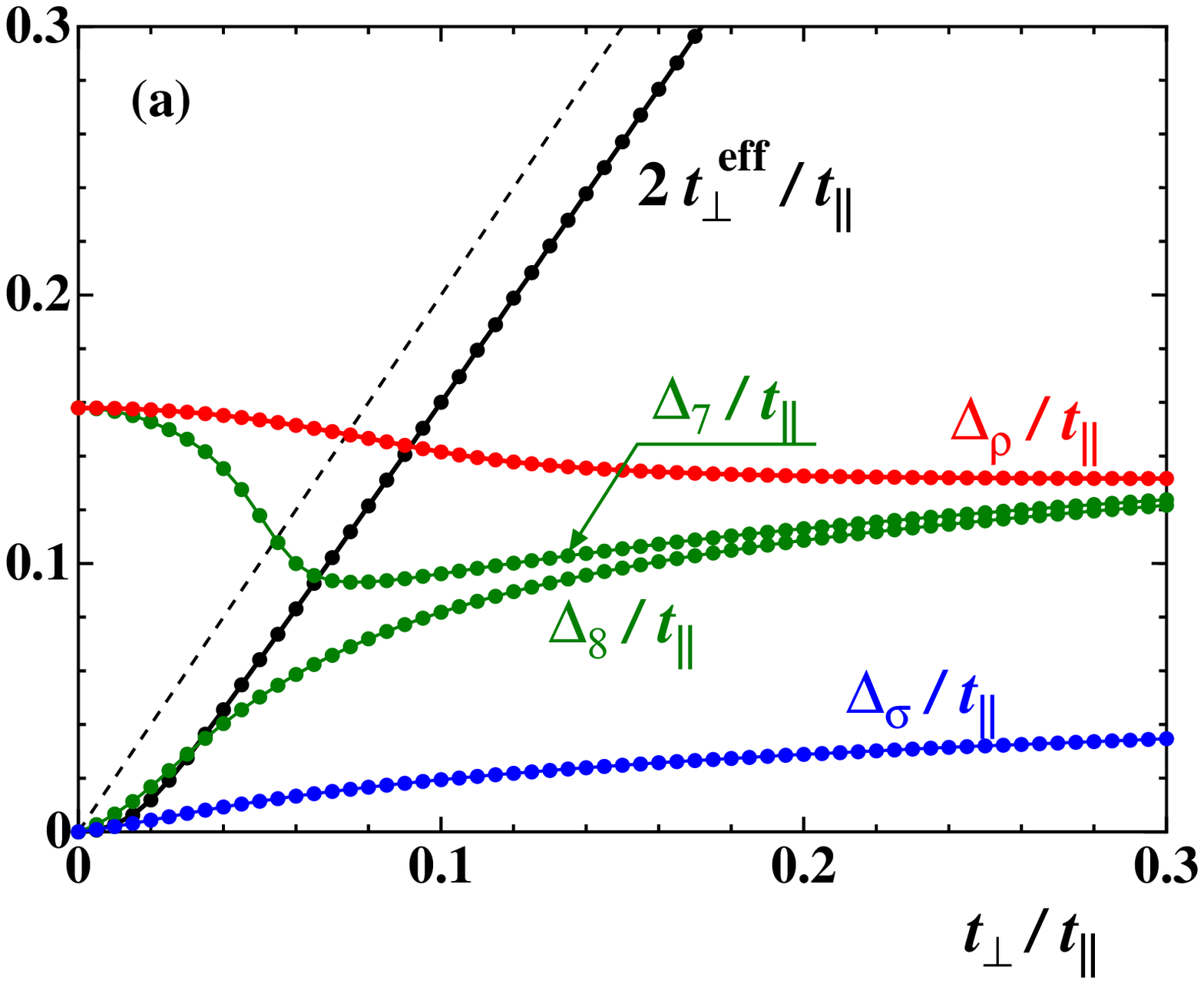}

\vspace*{.5cm}

\includegraphics[width=7cm]{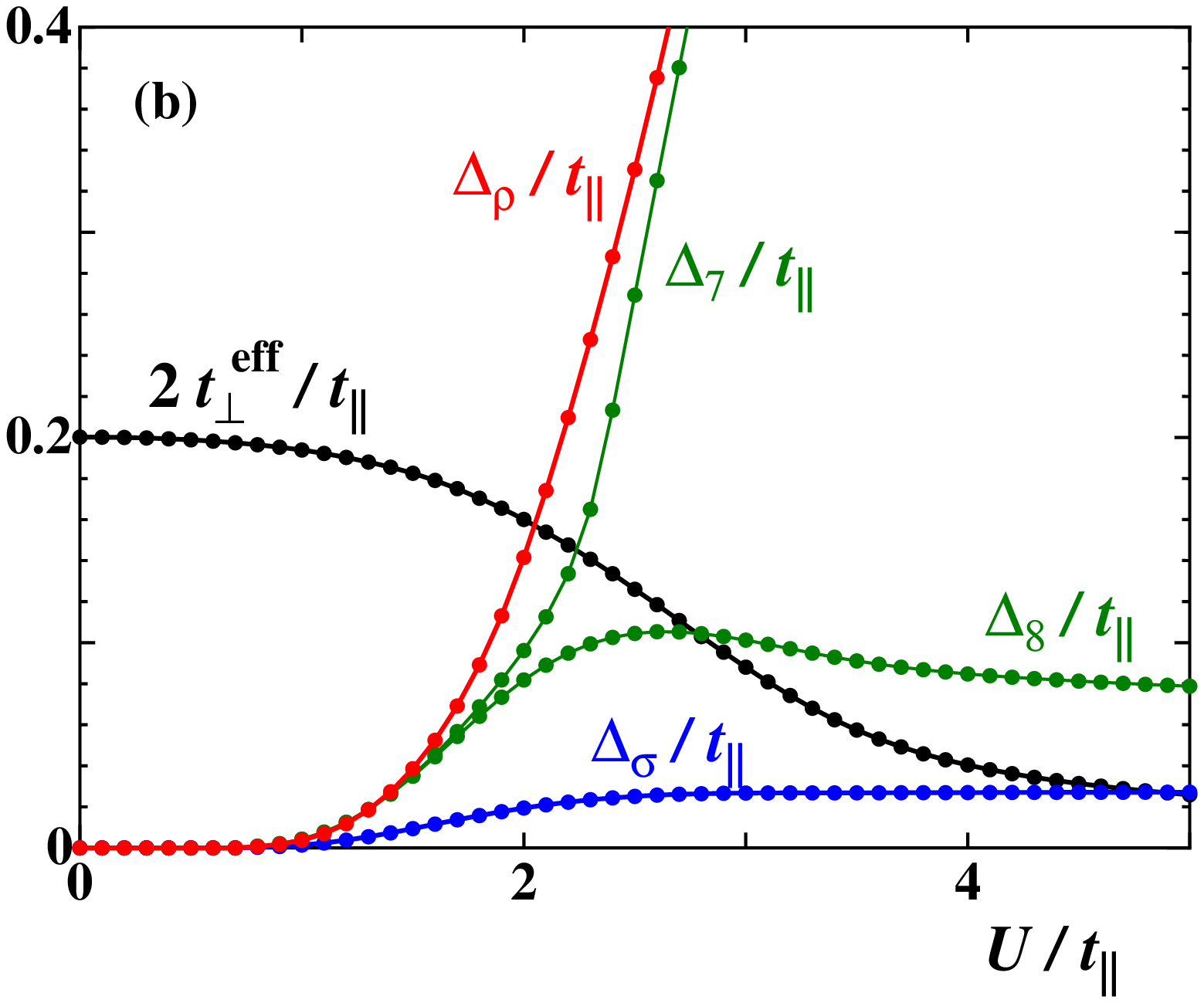}
\caption{
(Color online)
The excitation gaps, $\Delta_\rho$ (the charge gap),
  $\Delta_\sigma$ (the spin gap), 
  $\Delta_7$, $\Delta_8$ (the gaps in the Majorana fermion $\xi^7$ and
 $\xi^8$, see text), and the
  characteristic energy scale  $t_\perp^{\mathrm{eff}}$ for $N_\perp =2$. 
(a) The  $U/t_\parallel$ dependence with fixed 
  $t_\perp/t_\parallel=0.1$ and 
(b) the $t_\perp/t_\parallel$ dependence with fixed 
   $U/t_\parallel=2$.
The dashed line represents the magnitude of the bare interchain hopping.
}
\label{fig:Gap_ladder}
\end{figure}

The $U$ and $t_\perp$ dependences of 
 the charge and spin gaps and of the crossover 
  energy scale $t_\perp^{\mathrm{eff}}$ are shown 
  in Fig.\ \ref{fig:Gap_ladder}.
The $U/t_\parallel$ dependence of the spin gap  reproduce qualitatively 
  the DMRG numerical results, \cite{Noack}
  while our RG approach would  overestimate the magnitude of the spin gap.
As easily seen  from Fig.\ \ref{fig:Gap_ladder},
  the energy scales  of the charge and spin excitation gaps
  are different in the whole region of $U/t_\parallel$, which 
  is contrast to the analysis based on the one-loop fixed-point 
  behavior. \cite{Lin1998}

Next we examine the fixed-point behavior of the present analysis.
The fixed point values are
%====================================================================
\begin{eqnarray*}
&&
g_{\rho(\pi,0,0)}
=
g_{\rho(\pi,0,\pi)}
=
- g_{\sigma(0,0,0)}
=
g_{\sigma(\pi,0,\pi)}
= + g^*
,
\\ && 
g_{\rho(0,0,0)}
=
g_{\rho(0,0,\pi)}
= 
g_{\sigma(0,0,\pi)}
=
g_{\sigma(\pi,0,0)}
= 0,
\end{eqnarray*}
%====================================================================
where we find $g^*/(2\pi v) = 2$ in the present case.
This implies that 
  the symmetry is dynamically extended 
  \textit{in the low-energy limit}.
The effective theory in the low-energy limit has been analyzed 
  in the one-loop RG scheme and
  is known to be described as 
  the SO(8) Gross-Neveu model. \cite{Lin1998}
This effective theory can  easily be reproduced 
  from Eq.\ (\ref{eq:Hladder}).
To this end,
  we fermionize the $\phi_{\rho-}^p$ boson fields by introducing
 another set of  Majorana fermions: 
%===============================================================
\begin{subequations}
\begin{eqnarray}
 \frac{1}{\sqrt{2}} \left(\xi_{R}^8+i\xi_{R}^7\right)
 &\equiv& + \frac{\kappa_{\rho-}}{\sqrt{2\pi a}}
  \, \exp\left( i \, 2\phi_{\rho-}^R\right),
\\
\frac{1}{\sqrt{2}} \left(\xi_{L}^8+i\xi_{L}^7\right)
 &\equiv& - \frac{\kappa_{\rho-}}{\sqrt{2\pi a}}
  \, \exp\left( i \, 2\phi_{\rho-}^L\right),
\end{eqnarray}
\end{subequations}
%===============================================================
where $\kappa_{\rho -}$ is the Klein factor.
These Majorana fields satisfy the same anticommutation relations as before.
By using the Majorana fields   $\xi^n$ for $n=1,\cdots,8$ and
 by inserting the fixed-point values into Eq.\ (\ref{eq:Hladder}), 
  the fixed point Hamiltonian  can be expressed as
%===============================================================
\begin{eqnarray}
\mathcal{H}_{\mathrm{ladder}}^{\mathrm{eff}}
&=&
-i\frac{v}{2} 
 \sum_{n=1}^{8}
\left(
   \xi_{R}^n  \partial_x  \xi_{R}^n
-  \xi_{L}^n  \partial_x  \xi_{L}^n
\right)
\nonumber \\ && {}
+ \frac{g^*}{4}
\left(
\sum_{n=1}^{8}
\xi_{R}^n \, \xi_{L}^n 
\right)^2.
\end{eqnarray}
%===============================================================
 which is called the SO(8) Gross-Neveu model. \cite{Lin1998}
Here we note that this symmetry enlargement occurs 
\textit{in the low-energy limit}, where 
  all the excitations can be regarded to have
   the same magnitude of the excitation gap.
In the finite energy scale, however, this symmetry does not hold
  and has SO(3)$\times$SO(3)$\times$U(1) as seen in Eq.\ 
  (\ref{eq:Hladder}) for relevant interchain hopping.

%====================================================================
%======================================================================

Finally we examine the magnitude of the excitation 
  gaps for the remaining modes, $\xi^7$ and $\xi^8$, 
 and we show how the low-energy effective theory
   in the small interchain hopping 
  $t_\perp \ll \Delta_\rho$ can be described 
 and how the trivial limit of $t_\perp \to 0$
  can be reproduced in this Majorana-fermion description.
The form of the Hamiltonian (\ref{eq:Hladder}) is valid even in the
  small $t_\perp$ region, however, the physical picture 
   in the $t_\perp \to 0$ limit is not so trivial.
In terms of the Majorana fermions  $\xi^n$ ($n=1,\cdots,8$),
  the Hamiltonian (\ref{eq:Hladder}) in the $t_\perp \to 0$ limit
  can be rewritten as
   $\mathcal{H}_{\mathrm{eff}}|_{t_\perp \to 0}=\mathcal{H}_{\mathrm{eff}}^c
   +\mathcal{H}_{\mathrm{eff}}^s$ with
%===============================================================
\begin{subequations}
\begin{eqnarray}
\mathcal{H}_{\mathrm{eff}}^c
&=& 
-i\frac{v}{2} \sum_{n=4,5,6,7}
\left(
  \xi_R^n \partial_x \xi_{R}^n
- \xi_L^n \partial_x \xi_{L}^n
\right)
\nonumber \\ && {}
+ \frac{g_{\rho}}{2} 
\left( 
  \xi_{R}^4 \xi_{L}^4 + \xi_{R}^5 \xi_{L}^5
 +\xi_{R}^6 \xi_{L}^6 + \xi_{R}^7 \xi_{L}^7
\right)^2 ,
\\
\mathcal{H}_{\mathrm{eff}}^s
&=& 
-i\frac{v}{2} \sum_{n=1,2,3,8}
\left(
  \xi_R^n \partial_x \xi_{R}^n
- \xi_L^n \partial_x \xi_{L}^n
\right)
\nonumber \\ && {}
- \frac{g_{\sigma}}{2} 
\left( 
  \xi_{R}^1 \xi_{L}^1 + \xi_{R}^2 \xi_{L}^2
 +\xi_{R}^3 \xi_{L}^3 - \xi_{R}^8 \xi_{L}^8
\right)^2 , \qquad
\end{eqnarray}
\end{subequations}
%===============================================================
where $g_\rho$ becomes relevant and $g_\sigma$ becomes irrelevant.
Here we adopt  the notation $CnSm$ which denotes
  $n$ massless \textit{boson} modes in the charge sector and 
  $m$ massless \textit{boson} modes in the spin sector. \cite{Lin1997}
If one assigns that the bosonic phase variables $\phi_{\rho \pm}^p$ 
  and $\phi_{\sigma \pm}^p$ describe the ``charge'' and 
  ``spin'' modes respectively,
  the  $t_\perp \to 0$ limit may be interpreted as
  $C\frac{1}{2}S\frac{3}{2}$, where 
  the gapless ``spin'' mode is described by  the 
  $\bm \xi=(\xi^1,\xi^2,\xi^3)$ fermion
  (the central charge is $c=\frac{3}{2}$) and 
  the gapless ``charge'' mode is  by the $\xi^8$ fermion 
  (the central charge is $c=\frac{1}{2}$).
The total central charge is consistent with that for two isolated 
  Mott insulating chains $c=2$, however, this picture is not 
  correct obviously.
The correct understanding in the $t_\perp \to 0$ limit is that 
  the low-energy state is described by $C0S2$ where the Majorana fermions
  $\xi^7$  and $\xi^8$ should be regarded to describe the charge and spin 
  degrees of freedom respectively.
From this interpretation, we can expect that
  the magnitude of the gap in the Majorana fermions $\xi^7$ and $\xi^8$ 
  shows nontrivial behavior as a function of $t_\perp$, 
  since one ($\xi^7$) is gapped and 
  the other ($\xi^8$) is gapless in the $t_\perp\to 0$, 
  while these form the multiplet 
   and are transformed into the U(1) bosonic field $\theta_{\rho-}$
  in the large interchain hopping.
In order to estimate the $t_\perp$ dependence of the 
   gap in the Majorana fermions $\xi^7$ and $\xi^8$ of Eq.\ (\ref{eq:Hladder})
  from the numerical integration of the RG equations, 
 we consider the following combination of the coupling
%===============================================================
\begin{subequations}
\begin{eqnarray}
g_7
&=&
\frac{1}{2} 
 [g_{\rho(\pi,0,\pi)}+g_{\rho(0,0,\pi)}J(8t_\perp a/v)
\nonumber \\ && {}
+  g_{\sigma(\pi,0,\pi)}-g_{\sigma(0,0,\pi)} J(8t_\perp a/v)],
\\
g_8
&=&
\frac{1}{2}
 [g_{\rho(\pi,0,\pi)}-g_{\rho(0,0,\pi)}J(8t_\perp a/v)
\nonumber \\ && {}
+ g_{\sigma(\pi,0,\pi)}+g_{\sigma(0,0,\pi)}J(8t_\perp a/v)],
\end{eqnarray}
\end{subequations}
%===============================================================
where $J(8t_\perp a/v)$ is a cutoff function 
  satisfying $J(x)\approx 1$ for $x\ll 1$ and $J(x)\approx 0$ for $x\gg
  1$.
For relevant interchain hopping, we have
$g_7 =  g_8 = \frac{1}{2}  (g_{\sigma(\pi,0,\pi)} +g_{\rho(\pi,0,\pi)})$,
  which would reflect the low-energy property of the $\theta_{\rho-}$ 
  boson mode, and for $t_\perp\to 0$ we have
  $g_7 \to g_\rho$ and $g_8 \to g_\sigma$ reproducing the 
  single-chain limit.
The excitation gaps for these Majorana fermion  
  are also shown in Fig.\ \ref{fig:Gap_ladder}, where 
  we have estimated by $\Delta_{n}=\Lambda e^{-l_n}$ with 
  $G_n(l_n)=0.7$ ($n=7,8$).
The ground state of the present two-leg ladder system is 
  known to be the $D$-Mott phase for arbitrary $t_\perp>0$, however,
 as seen from Fig.\ \ref{fig:Gap_ladder}(a),
  the crossover from the 1D-like Mott insulating state
  (having large charge gap and small spin gap)  
 to the insulator of the ladder, which has 
  SO(3)$\times$SO(3)$\times$U(1) symmetry, takes place  
  at $t_\perp \approx \Delta_\rho|_{t_\perp=0}$
  where and the excitation properties
  for the Majorana fermion $\xi^7$ and $\xi^8$
  undergo considerable changes.
By increasing $U/t_\parallel$ [Fig.\ \ref{fig:Gap_ladder}(b)],
  the effective interchain hopping $t_\perp^{\mathrm{eff}}$ 
  is suppressed extremely and
  the multiplet of the $\xi^7$ and $\xi^8$ splits into 
  two isolated Majorana modes where the low energy excitations are  
  described by the Majorana triplet $\bm{\xi}=(\xi^1,\xi^2,\xi^3)$
  as a lowest-energy mode
  and by the Majorana singlet  $\xi^8$
  as a second-lowest-energy mode.
This picture reproduces the low-energy properties of 
  the Heisenberg spin ladder systems. \cite{Shelton}

The present estimations of the excitation gaps 
  are also justified by noting that
  it reproduces the known quantum critical behavior obtained 
  in the extended Hubbard model including the 
  intersite Coulomb repulsion.
The detailed estimation of the extended Hubbard model is given
  in the Appendix \ref{sec:appendix_ladder}.

\section{Conclusions}\label{sec:summary}

In the present paper, we have derived the two-loop RG equations 
  for the half-filled bipartite Q1D Hubbard model with the 
  nonperturbative treatment of the interchain hopping, 
  based on the conventional Kadanoff-Wilson approach.
By considering finite number of 1D chains 
  we have treated the transverse momentum $k_\perp$ 
   as the patch index and have
   obtained the RG equations which can be 
  extremely  simplified reflecting the symmetry requirements 
  of the Hubbard model.
By solving these RG equations numerically, 
  we have estimated the magnitude of the charge and spin gaps 
 and clarified that the charge gap 
  is suppressed due to the  interchain hopping  
  but is always finite even when the interchain hopping 
  exceeds the magnitude of the charge gap.
In order to justify the present approach,
  we have  analyzed the RG scaling flows in
   the two-leg Hubbard case ($N_\perp=2$) in detail based on 
  the field-theoretical Majorana-fermion description 
 and have clarified that the 
 low-energy excitations have SO(3)$\times$SO(3)$\times$U(1)  symmetry 
  for large interchain hopping.

\acknowledgments
 
The author thanks C.\ Bourbonnais, Y.\ Suzumura, and Y.\ Fuseya 
for valuable discussions at early stage of the present work.
The author also thanks  T.\ Giamarchi, A.\ Furusaki, D.K.\ Campbell
   for useful discussions and comments.
The numerical calculations were carried out in part 
  on Altix3700 BX2 at YITP in Kyoto University.

\appendix

\section{Charge-spin duality relation}\label{sec:appendix_duality}

In this section,
we derive the pseudospin SU(2) relations (\ref{eq:chargeSU(2)_gology})
 from  the spin SU(2) relations (\ref{eq:spinSU(2)_gology})
  by using the ``charge-spin duality'' transformation.
It is well known that the  Hubbard Hamiltonian (\ref{eq:model})
  is transformed to itself with 
  $U\to -U$, under 
  the particle-hole transformation for the
  spin down only, \cite{Shiba,Nagaoka} i.e., 
%====================================================================
\begin{eqnarray}
  c_{j,l,\uparrow} \leftrightarrow c_{j,l,\uparrow}, \quad
  c_{j,l,\downarrow} \leftrightarrow 
  (-1)^{j+l} c_{j,l,\downarrow}^\dagger.
\label{eq:a1}
\end{eqnarray}
%====================================================================
Since the density operators are transformed as 
   $(n_{j,l,\uparrow}+n_{j,l,\downarrow}) \leftrightarrow
    (n_{j,l,\uparrow}-n_{j,l,\downarrow})$
  under this transformation,
   the charge and spin density operators are interchanged.
In the Fourier space with the linearized dispersion,
 Eq.\ (\ref{eq:a1}) is rewritten as
%====================================================================
\begin{equation}
c_{p,\uparrow}(\bm k)  \leftrightarrow  c_{p,\uparrow}  (\bm k) , \quad
c_{p,\downarrow}(\bm k) \leftrightarrow
c_{p,\downarrow}^\dagger  ((p\pi,\pi)-\bm k). 
\end{equation}
%====================================================================
By applying this transformation to the 
   $g$-ology Hamiltonian (\ref{eq:g-ology}), 
  we find that the transformed Hamiltonian is given 
   in the same form of Eq.\ (\ref{eq:g-ology}), but 
 the coupling constants are exchanged as
%====================================================================
\begin{subequations}
\begin{eqnarray}
g_{1\perp(q,k_1,k_2)}
&\leftrightarrow& - g_{3\perp(q,k_1,\pi-k_2)} ,  \\
g_{2\perp(q,k_1,k_2)}
&\leftrightarrow& -g_{2\perp(\pi-q+k_1+k_2,k_1,k_2)} ,
\end{eqnarray}%
\label{eq:duality_in_g}%
\end{subequations}
%====================================================================
while $g_{\parallel(q,k_1,k_2)}$ and $g_{3\parallel(q,k_1,k_2)}$
  are unchanged.
In the spin part there are constraints (\ref{eq:spinSU(2)_gology})
   due to the  spin-rotational SU(2)  symmetry.
By applying the duality relation (\ref{eq:duality_in_g})
  to Eq.\  (\ref{eq:spinSU(2)_gology}),
  we can derive the pseudospin SU(2)
   constraints (\ref{eq:chargeSU(2)_gology}).

\section{Full RG equations for the spin-rotational invariant case}
\label{sec:appendix_RG}

In this section, the full two-loop RG equations are given in the 
  case for the spin-rotational SU(2) symmetric case 
  [Eq.\ (\ref{eq:spinSU(2)})],
  without assuming the pseudospin SU(2) symmetry [Eq.\ (\ref{eq:chargeSU(2)})].
These RG equations are valid for the extended Hubbard model 
  including additional spin-rotational symmetric interactions,
  e.g., intersite Coulomb repulsions. 

The RG equation for the interchain hopping is given by
%====================================================================
\begin{eqnarray}
\frac{d}{dl}
t_{\perp} &=&
t_{\perp}
-\frac{1}{4N_\perp^3} \sum_{q ,k, k'}
G_{\Sigma \mathrm{n}(q,k,k')}^2 \, J_{0(q,k,k')} \cos k
\nonumber \\ && {}
-\frac{1}{4N_\perp^3} \sum_{q ,k, k'}
G_{\Sigma \mathrm{u}(q,k,k')}^2 \, J'_{0(q,k,k')} \cos k,
\label{eq:RG_tperp_app}
\end{eqnarray}
%====================================================================
where the second-order coupling constants contributing the
  self-energy corrections are put into forms:
%====================================================================
\begin{subequations}
\begin{eqnarray}
G_{\Sigma \mathrm{n}(q,k,k')}^2
&\equiv&
\frac{1}{2}
\Bigl[
  G_{\rho(q,k,k')}^2 
+ 3\, G_{\sigma(q,k,k')}^2
\Bigr],
\\
G_{\Sigma \mathrm{u}(q,k,k')}^2
&\equiv&
\frac{1}{2}
\Bigl[
  2 \, G_{c(q,k,k')}^2
+ 2 \, G_{c(\pi-q+k+k',k,k')}^2
\nonumber \\ && {}
- 2 \, G_{c(q,k,k')}G_{c(\pi-q+k+k',k,k')}
\Bigr].
\end{eqnarray}
\end{subequations}
%====================================================================
The cutoff function $J_{0(q,k,k')}$ is given by Eq.\ (\ref{eq:J0}).
In general, the cutoff function for the umklapp scattering contributions
  $J'_{0(q,k,k')}$ takes a different form from that for the normal scattering
 ones $J_{0(q,k,k')}$, however, 
  if the system has the particle-hole symmetry, these become identical
  $J_{0(q,k,k')}=J'_{0(q,k,k')}$.

The RG equations for the coupling constants 
  without the assumption of the spin-rotational SU(2) symmetry are 
  given in symbolic form as
\begin{widetext}
%====================================================================
\begin{subequations}
\begin{eqnarray}
\frac{d}{dl}
G_{\rho(q,k_1,k_2)}
&=&
\left[
\frac{1}{2N_\perp} \sum_{k'}
\Xi_{\rho1(q,k_1,k_2,k')}
+ \frac{1}{8N_\perp^2} \sum_{q',k'}
\Xi_{\rho2(q,k_1,k_2,q',k')}
+ (k_1 \leftrightarrow k_2)
\right]
\nonumber \\ && {}
- \frac{1}{4N_\perp^2}  \,  G_{\rho(q,k_1,k_2)} 
\sum_{q' ,k'}   \Xi_{3(q,k_1,k_2,q',k')},
\label{eq:RG_Grho}
\\
\frac{d}{dl} G_{\sigma(q, k_1,k_2)} 
&=& 
\left[
 \frac{1}{2N_\perp} \sum_{k'}
\Xi_{\sigma1(q,k_1,k_2,k')}
+ \frac{1}{8N_\perp^2} \sum_{q',k'}
\Xi_{\sigma2(q,k_1,k_2,q',k')}
+ (k_1 \leftrightarrow k_2)
\right]
\nonumber \\ && {}
- \frac{1}{4N_\perp^2}  \,  G_{\sigma(q,k_1,k_2)} 
\sum_{q' ,k'}  \Xi_{3(q,k_1,k_2,q',k')}
,
\label{eq:RG_Gsigma}
\\
\frac{d}{dl}
G_{c(q,k_1,k_2)} 
&=&
\left[
 \frac{1}{2N_\perp} \sum_{k'}
\Xi_{c1(q,k_1,k_2,k')}
+ \frac{1}{8N_\perp^2} \sum_{q',k'}
\Xi_{c2(q,k_1,k_2,q',k')}
+\Bigr((q,k_1,k_2)\to (-q,\pi-k_2,\pi-k_1)\Bigr)
\right]
\nonumber \\ && {}
- \frac{1}{4N_\perp^2}  \,  G_{c(q,k_1,k_2)} 
\sum_{q' ,k'}    \Xi_{c3(q,k_1,k_2,q',k')},
\label{eq:RG_Gc}
\end{eqnarray}
\end{subequations}
%====================================================================
where $\Xi_{\nu 1}$ and $\Xi_{\nu 2}$
  represent the one-loop Peierls/Cooper bubble contributions 
  and the two-loop (third-order) vertex contributions, respectively, 
  and $\Xi_{3}$ and $\Xi_{c3}$ represent 
  the two-loop (second-order) self-energy contributions.
The respective terms are given explicitly in the following.
The one-loop Peierls and Cooper bubble contributions 
  are given by
%====================================================================
\begin{subequations}
\begin{eqnarray}
\Xi_{\rho1(q,k_1,k_2,k')}
&=&
\frac{1}{2}
\bigl[
    G_{\rho(q,k_1,k')} \,  G_{\rho(q,k',k_2)} 
  +3G_{\sigma(q,k_1,k')} \,  G_{\sigma(q,k',k_2)} 
\bigr]
  I_{(q,k',k_1,k_2)}
\nonumber \\  && {}
- \frac{1}{2}
\bigl[
      G_{\rho(q-k_2+k',k_1,k')} \, 
      G_{\rho(q-k_1+k', k',k_2)} \,
  + 3 G_{\sigma(q-k_2+k',k_1,k')} \, 
      G_{\sigma(q-k_1+k', k',k_2)} 
\bigr]
 I_{C(q-k_1-k_2 ,k',k_1,k_2)}
\nonumber \\ && {}
+ 
2 \bigl[
    G_{c(q,k_1,k')} \,  G_{c(q,k_2,k')}
  + G_{c(\pi-q+k_1+k',k_1,k')} \, G_{c(\pi+q-k_2-k',\pi-k',\pi-k_2)} 
\nonumber \\ && {} \qquad
  - G_{c(q,k_1,k')} \,  G_{c(\pi+q-k_2-k',\pi-k',\pi-k_2)}
\bigr]
  I'_{(-q,\pi-k',\pi-k_1,\pi-k_2)} ,
\\ %--------
\Xi_{\sigma1(q, k_1,k_2,k')}
&=& 
\bigl[
    G_{\rho(q,k_1,k')} \,  G_{\sigma(q,k',k_2)} 
  - G_{\sigma(q,k_1,k')} \,  G_{\sigma(q,k',k_2)}
\bigr]
  I_{(q,k',k_1,k_2)}
\nonumber \\  && {}
- 
\bigl[
  G_{\rho(q-k_2+k',k_1,k')}
 +G_{\sigma(q-k_2+k',k_1,k')} 
\bigr] G_{\sigma(q-k_1+k', k',k_2)}
 I_{C(q-k_1-k_2 ,k',k_1,k_2)}
\nonumber \\  && {}
- 2 \bigl[
    G_{c(q,k_1,k')} \,  G_{c(q,k_2,k')} 
 -  G_{c(q,k_1,k')} \,  G_{c(\pi+q-k_2-k',\pi-k',\pi-k_2)}
\bigr]
  I'_{(-q,\pi-k',\pi-k_1,\pi-k_2)}  ,
\\ %--------
\Xi_{c1(q,k_1,k_2,k')} 
&=&
\bigl[
  G_{\rho(q,k_1,k')} G_{c(q,k',k_2)} 
 - 3 G_{\sigma(q,k_1,k')} \, G_{c(q,k',k_2)} \,
+ 2G_{\sigma(q,k_1,k')} \,
  G_{c(\pi-q+k'+k_2,k',k_2)} \,
\bigr]
  I''_{(q,k',k_1,k_2)}
\nonumber \\ &&{}
+ 
\bigl[
  G_{\rho(\pi-q+k_1+k_2,k_1,k')} \, G_{c(q-k_1+k',k',k_2)}
  + G_{\sigma(\pi-q+k_1+k_2,k_1,k')} \, G_{c(q-k_1+k',k',k_2)}
\bigr]
  I''_{(\pi-q+k_1+k_2,k',k_1,k_2)}  .
\nonumber \\ 
\end{eqnarray}
\end{subequations}
%====================================================================
Due to the particle-hole symmetry
   of the present model on the bipartite lattice,
  the respective cutoff
  functions satisfy $I'_{(q,k',k_1,k_2)}=I''_{(q,k',k_1,k_2)}
  =I_{(q,k',k_1,k_2)}$,
  $I_{(-q,\pi-k',\pi-k_1,\pi-k_2)} = I_{(q,k',k_1,k_2)}$, and
  $I_{C(q-k_1-k_2,k',k_1,k_2)} = I_{(\pi-q+k_1+k_2,k',k_1,k_2)}$,
  where $I_{(q,k',k_1,k_2)}$ is given in Eq.\ (\ref{eq:I}).
The two-loop vertex contributions are given by
%====================================================================
\begin{subequations}
\begin{eqnarray}
\Xi_{\rho2(q,k_1,k_2,q',k')}
&=&
G_{\rho(q+q',k_1,k_2)} 
\bigl[
      G_{\rho(q-k_1+k',k',k'-q')}  \, 
      G_{\rho(q-k_2+k',k',k'-q')} 
\nonumber \\ && {} \qquad
  + 3 G_{\sigma(q-k_1+k',k',k'-q')}  \,
      G_{\sigma(q-k_2+k',k',k'-q')} 
\bigr]
     J_{2(q+k';k_1,k_2;k',k'-q')}
\nonumber \\ && {} 
- 2  G_{\rho(\pi-q-q'+k_1+k_2,k_1,k_2)}
\bigl[
   G_{c(q-k_1+k',k',k'-q')} \,  G_{c(q-k_2+k',k',k'-q')} 
\nonumber \\ && {} \qquad
+  G_{c(\pi-q-q'+k_1+k',k',k'-q')} \,
   G_{c(\pi-q-q'+k_2+k',k',k'-q')} 
\nonumber \\ && {} \qquad
-  G_{c(q-k_1+k',k',k'-q')} \,
   G_{c(\pi-q-q'+k_2+k',k',k'-q')} 
\bigr]
  J'_{2(q+k';k_1,k_2;k',k'-q')}
\nonumber \\ && {}
+ G_{\rho(q-q',k_1-q',k_2-q')}
  \bigl[
  G_{\rho(k_1-k',k_1,k_1-q')} \,
  G_{\rho(k_2-k',k_2,k_2-q')}
\nonumber \\ && {}  \qquad
+ 3   G_{\sigma(k_1-k',k_1,k_1-q')} \,
      G_{\sigma(k_2-k',k_2,k_2-q')} 
\bigr]
  J_{2(-k';-k_1,-k_2;\pi-k',\pi-k'+q')}
\nonumber \\ && {} 
- 2 G_{\rho(\pi+q+q'-k_1-k_2,\pi-k_1+q',\pi-k_2+q')} 
\bigl[
    G_{c(k_1-k',k_1,k_1-q')} \, 
    G_{c(k_2-k',k_2,k_2-q')}  
\nonumber \\ && {} \qquad
+   G_{c(\pi-q'+k_1+k',k_1,k_1-q')} \,
    G_{c(\pi-q'+k_2+k',k_2,k_2-q')}
\nonumber \\ && {} \qquad
-   G_{c(k_1-k',k_1,k_1-q')} \,
    G_{c(\pi-q'+k_2+k',k_2,k_2-q')} 
\bigr]
  J'_{2(-k';-k_1,-k_2;\pi-k',\pi-k'+q')}
,
\\ 
\Xi_{\sigma2(q,k_1,k_2,q',k')}
&=&
G_{\sigma(q+q',k_1,k_2)}
\bigl[
  G_{\rho(q-k_1+k',k',k'-q')}  \,
  G_{\rho(q-k_2+k',k',k'-q')} 
\nonumber \\ && {} \qquad
 - G_{\sigma(q-k_1+k',k',k'-q')}  \,
   G_{\sigma(q-k_2+k',k',k'-q')}
\bigr]
  J_{2(q+k';k_1,k_2;k',k'-q')}
\nonumber \\ && {} 
+ 2
  G_{\sigma(\pi-q-q'+k_1+k_2,k_1,k_2)} \, 
  G_{c(q-k_1+k',k',k'-q')} \,
  G_{c(\pi-q-q'+k_2+k',k',k'-q')}  \,
  J'_{2(q+k';k_1,k_2;k',k'-q')}
\nonumber \\ && {}
+ G_{\sigma(q-q',k_1-q',k_2-q')}
 \bigl[
    G_{\rho(k_1-k',k_1,k_1-q')} \,
    G_{\rho(k_2-k',k_2,k_2-q')} 
\nonumber \\ && {}\qquad
  - G_{\sigma(k_1-k',k_1,k_1-q')} \,
    G_{\sigma(k_2-k',k_2,k_2-q')} 
\bigr]
  J_{2(-k';-k_1,-k_2;\pi-k',\pi-k'+q')}
\nonumber \\ && {} 
+ 2
  G_{\sigma(\pi+q+q'-k_1-k_2,\pi-k_2+q',\pi-k_1+q')} \, 
  G_{c(k_1-k',k_1,k_1-q')} \,
  G_{c(\pi-q'+k_2+k',k_2,k_2-q')}  \,
\nonumber \\ && {} \qquad \times
  J'_{2(-k';-k_1,-k_2;\pi-k',\pi-k'+q')}
,
\\ 
\Xi_{c2(q,k_1,k_2,q',k')}
&=&
\bigl[
  2G_{c(\pi-q-q'+k_1+k_2,k_1,k_2)}\,
  G_{\sigma(q-k_1+k',k',k'-q')} \, 
  G_{\sigma(\pi-q-q'+k_2+k',k',k'-q')}
\nonumber \\ && {} \qquad
-  G_{c(q+q',k_1,k_2)}\,
   G_{\rho(q-k_1+k',k',k'-q')} \, 
   G_{\rho(\pi-q-q'+k_2+k',k',k'-q')} 
\nonumber \\ && {} \qquad
+  G_{c(q+q',k_1,k_2)}\,
   G_{\sigma(q-k_1+k',k',k'-q')} \, 
   G_{\sigma(\pi-q-q'+k_2+k',k',k'-q')}
\bigr]
  J''_{2(q+k';k_1,k_2;k',k'-q')}
\nonumber \\ && {}
+ 
\bigl[
   2 G_{c(\pi-q-q'+k_1+k_2,k_1-q',k_2-q')}\,
     G_{\sigma(k_1-k',k_1,k_1-q')} \, 
     G_{\sigma(\pi+q'-k_2-k',\pi-k_2,\pi-k_2+q')} \,
\nonumber \\ && {} \qquad
- G_{c(q-q',k_1-q',k_2-q')}\,
  G_{\rho(k_1-k',k_1,k_1-q')} \, 
  G_{\rho(\pi+q'-k_2-k',\pi-k_2,\pi-k_2+q')} 
\nonumber \\ && {} \qquad
+ G_{c(q-q',k_1-q',k_2-q')}\,
  G_{\sigma(k_1-k',k_1,k_1-q')} \, 
  G_{\sigma(\pi+q'-k_2-k',\pi-k_2,\pi-k_2+q')}
\bigr]
\nonumber \\ && {} \qquad\times
  J''_{2(-k';-k_1,-k_2;\pi-k',\pi-k'+q')}
,
\end{eqnarray}
\end{subequations}
%====================================================================
and the two-loop self-energy contributions  are given by
%====================================================================
\begin{subequations}
\begin{eqnarray}
 \Xi_{3(q,k_1,k_2,q',k')}
&=&
  G_{\Sigma \mathrm{n}(q',k_1,k')}^2 \, J_{1(q',k_1,k')}
+ G_{\Sigma \mathrm{u}(q',k_1,k')}^2 \, J'_{1(q',k_1,k')}
\nonumber \\ && {} 
+ G_{\Sigma \mathrm{n}(q',k_2,k')}^2 \, J_{1(q',k_2,k')}
+ G_{\Sigma \mathrm{u}(q',k_2,k')}^2 \, J'_{1(q',k_2,k')}
\nonumber \\ && {} 
+ G_{\Sigma \mathrm{n}(q',-k_1+q,k')}^2 \, J_{1(q',-k_1+q,k')}
+ G_{\Sigma \mathrm{u}(q',-k_1+q,k')}^2 \, J'_{1(q',-k_1+q,k')}
\nonumber \\ && {} 
+ G_{\Sigma \mathrm{n}(q',-k_2+q,k')}^2 \, J_{1(q',-k_2+q,k')}
+ G_{\Sigma \mathrm{u}(q',-k_2+q,k')}^2 \, J'_{1(q',-k_2+q,k')},
\\
 \Xi_{c3(q,k_1,k_2,q',k')}
&=&
  G_{\Sigma \mathrm{n}(q',k_1,k')}^2 \, J_{1(q',k_1,k')}
+ G_{\Sigma \mathrm{u}(q',k_1,k')}^2 \, J'_{1(q',k_1,k')}
\nonumber \\  && {}
+ G_{\Sigma \mathrm{n}(q',\pi-k_2,k')}^2 \, J_{1(q',\pi-k_2,k')}
+ G_{\Sigma \mathrm{u}(q',\pi-k_2,k')}^2 \, J'_{1(q',\pi-k_2,k')}
\nonumber \\  && {}
+ G_{\Sigma \mathrm{n}(q',-k_1+q,k')}^2 \, J_{1(q',-k_1+q,k')}
+ G_{\Sigma \mathrm{u}(q',-k_1+q,k')}^2 \, J'_{1(q',-k_1+q,k')}
\nonumber \\  && {}
+ G_{\Sigma \mathrm{n}(q',\pi+k_2-q,k')}^2 \, J_{1(q',\pi+k_2-q,k')}
+ G_{\Sigma \mathrm{u}(q',\pi+k_2-q,k')}^2 \, J'_{1(q',\pi+k_2-q,k')}.
\end{eqnarray}
\end{subequations}
%====================================================================
\end{widetext}
The cutoff functions  $J_1$ and $J_1'$ 
  ($J_2$, $J_2'$, and $J_2''$) depend on the lattice geometry of the 
  model and  take different forms in general. 
However, in the present bipartite model, the respective cutoff
 functions satisfy 
  $J'_{1(q,k,k')}=J_{1(q,k,k')}$,
  $J'_{2(q;k_1,k_2;k',k'')}=J''_{2(q;k_1,k_2;k',k'')}
                           =J_{2(q;k_1,k_2;k',k'')}$.
We also obtain   
  $J_{1(-q,\pi-k_1,\pi-k_2)}=J_{1(q,k_1,k_2)}$ and
  $J_{2(-k';-k_1,-k_2;\pi-k',\pi-k'+q')} = J_{2(k';k_1,k_2;k',k'-q')}$
  for the particle-hole symmetric case.

If the interaction is on-site one only,
  the system has the pseudospin SU(2)  symmetry, where
  Eq.\ (\ref{eq:chargeSU(2)_gology}) is satisfied.
By using Eq.\ (\ref{eq:chargeSU(2)_gology}), the coupling constant
  $G_{\Sigma(q,k,k')}^2=G_{\Sigma \mathrm{n}(q,k,k')}^2
                       +G_{\Sigma \mathrm{u}(q,k,k')}^2$
  can be rewritten in terms of $G_\rho$ and $G_\sigma$ 
  and reproduces Eq.\ (\ref{eq:Gself}).
Then the RG equation for the interchain 
  hopping [Eq.\ (\ref{eq:RG_tperp_app})] leads Eq.\ (\ref{eq:RG_tperp})
  and those for the coupling constants
  [Eqs.\ (\ref{eq:RG_Grho}) and (\ref{eq:RG_Gsigma})]
  lead Eq.\ (\ref{eq:RG_G}).
The explicit RG equations for the umklapp scattering [Eq.\ (\ref{eq:RG_Gc})]
  can be suppressed due to the pseudospin SU(2) symmetry 
  [Eq.\ (\ref{eq:chargeSU(2)_gology})].

\section{Extended Two-Leg Ladder Model: 
Check of quantum critical behavior}
\label{sec:appendix_ladder}

In Sec.\ \ref{sec:formulation}, we have estimated the magnitudes of 
  charge and spin excitation gaps by using  Eq.\ (\ref{eq:gapestimation}).
If the charge and spin modes of the system are decoupled, 
  such as in the single chain case,
  this method trivially works since the coupling constants
   representing respective modes are decoupled.
However in the present $N_\perp$-chain system where
  the charge and spin degrees of freedom coupled with each other,
  one may consider that the present analysis does not work 
  since the RG approach may break down at a energy scale 
  corresponding to the largest excitation gap.
In order to justify the present estimation of 
  excitation gaps, we have considered 
  the two-leg ladder model ($N_\perp=2$) which is a minimal 
  model with the spin and charge modes coupled.
As already mentioned in Sec.\ \ref{sec:ladder},
  the $U$ dependence of the spin gap [Fig.\ \ref{fig:Gap_ladder} (b)]
  shows similar behavior to the DMRG results  \cite{Noack}.
In this section, we reconsider the two-leg ladder systems and
  we show another evidence which supports strongly 
  the validity of the present estimation of excitation gaps.

We consider a toy model including an additional interaction $V'$
  which denotes the next-nearest-neighbor Coulomb repulsion.
The spin mode in this model is known to exhibit
   quantum critical behavior within 
  a nontrivial universality class.
The purpose of the present section is to 
  check whether the present method 
  reproduces correct behavior of the quantum critical point (QCP).
The Hamiltonian of this toy model is given by
%====================================================================
\begin{eqnarray}
H' &=& 
-t_\parallel \sum_{j,l,s} 
\left(
  c_{j,1,s}^\dagger c_{j+1,1,s}
+ c_{j,2,s}^\dagger c_{j+1,2,s}
+\mathrm{H.c.} 
\right)
\nonumber \\ && {}
-2t_\perp \sum_{j,s} 
\left(c_{j,1,s}^\dagger c_{j,2,s}+\mathrm{H.c.} \right)
\nonumber \\ && {}
+ U \sum_{j}
\left(
 n_{j,1,\uparrow} n_{j,1,\downarrow} 
+ n_{j,2,\uparrow} n_{j,2,\downarrow} 
\right)
\nonumber \\ && {}
+
V' \sum_j
\left(
n_{j,1} n_{j+1,2} + n_{j,2} n_{j+1,1}
\right).
\end{eqnarray}
%====================================================================
The notations are the same as in Eq.\ (\ref{eq:model}). 
This extended two-leg ladder model is examined by
   the field-theoretical method \cite{Tsuchiizu2002}.
For small $V'$,  the rung-singlet (or $D$-Mott) state is realized where
  the ground state is unique.
By increasing  $V'$, this rung-singlet state changes into 
  a spin-Peierls (or PDW) state 
  (see Fig.\ 9 of Ref.\ \onlinecite{Tsuchiizu2002})
  where the ground state has two-fold degeneracy 
  and breaks translational invariance along the chain direction.
From the field-theoretical approach,
  the  quantum critical behavior is confirmed on the transition
  point between the rung-singlet state and the spin-Peierls state. 
On this QCP, the spin gap collapses 
  and the effective theory for low-energy states is known to be 
  described by the $c=3/2$ conformal field theory where $c$ is the
  central charge.

\begin{figure}[t]
\includegraphics[width=7cm]{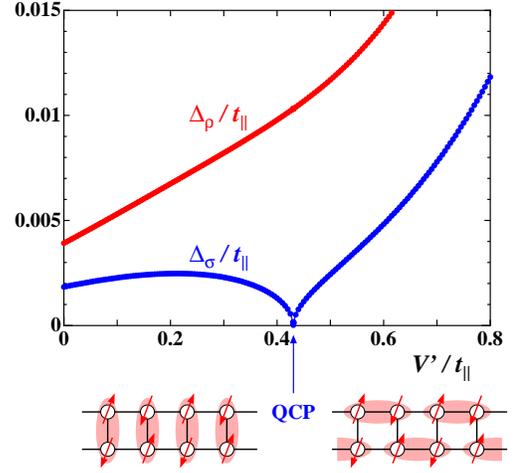}
\caption{
(Color online)
The $V'/t_\parallel$ dependences of
the charge gap $\Delta_\rho$ and the spin gap $\Delta_\sigma$,
  for the extended two-leg ladder model ($N_\perp=2$) with
  $U/t_\parallel=1$ and $t_\perp/t_\parallel=0.5$.
}
\label{fig:Gap_ladderV}
\end{figure}

This extended Hubbard model can also be analyzed in the present
  framework of the two-loop RG, where the only differences 
  from the analysis in Sec.\ \ref{sec:formulation} are that
  (i) the $g$-ology coupling constants in Eq.\ (\ref{eq:g-ology}) 
  have explicit momentum dependence and (ii)
  the pseudospin SU(2) [Eq.\ (\ref{eq:chargeSU(2)_gology})]
  is not retained due to the presence of the additional interaction.
The RG equations in this generalized case are given in the Appendix
  \ref{sec:appendix_RG}.
The estimated charge and spin gaps as a function of $V'/t_\parallel$
  is shown in Fig.\ \ref{fig:Gap_ladderV}.
We find that the present approach reproduces the 
  critical behavior since the spin gap becomes small around the QCP and 
  collapses just on the QCP.
The critical value of $V'$ is 
  consistent with Fig.\ 9 in Ref.\ \onlinecite{Tsuchiizu2002}.
The RG scaling flows on the QCP show that
  the coupling $G_{\rho+}$ reaches of the order unity 
  for $l>l_{\rho+}$, however, the coupling $|G_{\sigma+}|$ remains
  small and becomes irrelevant $G_{\sigma+}(\infty)=0$.
Such scaling behavior is the same as expected 
  from the field-theoretical approach, \cite{Tsuchiizu2002}
  and thus the present results 
  can be justified even for spin-charge coupled systems.

%====================================================================
%======================================================================

From the technical point of view, we discuss the reason why the 
  present analysis works even for spin-charge coupled systems.
If one of the coupling constants reaches of the order unity 
  in the scaling flow, 
  the RG method breaks down where it can be understood that 
  the corresponding mode has an excitation gap.
In order to analyze the lower-energy properties further,  
  the gapped mode should be traced out and 
  one should derive the effective low-energy theory
  for remaining modes. 
Then one can apply the RG method to it again. 
In this context, the quantum critical behavior 
  was confirmed in Refs.\ \onlinecite{Tsuchiizu2002} and 
  \onlinecite{Fradkin2003}.
As for the two-leg ladder systems, we find that
this tracing-out procedure almost corresponds to the replacement of 
the relevant coupling constants to unity.
On the other hand, 
 in the scaling flow of the present two-loop RG, the coupling constants
 remain finite even for the relevant ones (see Fig.\ \ref{fig:flow8}).
Thus one can consider that such trancing-out procedure of the gapped mode is
performed automatically in the present two-loop RG approach.
The minor differences between these two approaches 
  do not affect the numerical results.
Thus we find that the present approach to estimate the different energy
  gaps works even for spin-charge coupled systems.

Finally we note that
  the procedure of the derivation of the effective theory
  is not straightforward and restricted to the $N_\perp=2$ case only.
In the present estimation based on the two-loop RG,
  there is no need to derive such low-energy effective theory 
  explicitly and thus 
this fact is the reason why it is easy to extend the analysis 
  to the large number of chains systems.


\begin{thebibliography}{}

%--------------------------
\bibitem{Bourbonnais2003}
C.\ Bourbonnais, B.\ Guay, and R.\ Wortis,
in \textit{Theoretical Methods for Strongly Correlated Electrons}
edited by D.\ S\'en\'echal, A.M.\ Tremblay, and C.\ Bourbonnais
(Springer, New York, 2003), p.\ 77.
%--------------------------
%--------------------------
\bibitem{Emery}
V.J.\ Emery, in
\textit{Highly Conducting One-Dimensional Solids},
  edited by J.\ Devreese, R.\ Evrard, and V.\ van Doren
  (Plenum, New York, 1979), p.\ 247.
%--------------------------
%--------------------------
\bibitem{Solyom}
J.\ S\'olyom,
Adv.\ Phys.\ \textbf{28}, 201 (1979).
%--------------------------
%--------------------------
\bibitem{Bourbonnais1991}
C.\ Bourbonnais and L.G.\ Caron,
Int.\ J.\ Mod.\ Phys.\ B \textbf{5}, 1033 (1991).
%--------------------------
%--------------------------
\bibitem{Giamarchi_book}
T. Giamarchi,
\textit{Quantum Physics in One Dimension}
(Oxford University Press, 2004).
%--------------------------
%--------------------------
\bibitem{Shankar}
R.\ Shankar,
Rev. Mod. Phys. \textbf{66}, 129 (1994).
%--------------------------
%--------------------------
\bibitem{Furukawa} 
N.\ Furukawa, T.M.\ Rice, and M.\ Salmhofer,
Phys.\ Rev.\ Lett. \textbf{81}, 3195 (1998), and references therein.
%--------------------------
%--------------------------
\bibitem{Zanchi1998}
D.\ Zanchi and H.J.\ Schulz,
Europhys.\ Lett.\  \textbf{44}, 235 (1998);
Phys.\ Rev.\ B \textbf{61}, 13609 (2000).
%--------------------------
%--------------------------
\bibitem{Salmhofer1998}
M.\ Salmhofer,
Commun.\ Math.\ Phys.\ \textbf{194}, 249 (1998).
%--------------------------
%--------------------------
\bibitem{Halboth2000}
C.J.\ Halboth and W.\ Metzner,
Phys.\ Rev.\ B \textbf{61}, 7364 (2000).
%--------------------------
%--------------------------
\bibitem{Honerkamp2001}
C.\ Honerkamp, M.\ Salmhofer, N.\ Furukawa, and T.M.\ Rice,
Phys.\ Rev.\ B \textbf{63}, 035109 (2001);
M.\ Salmhofer and C.\ Honerkamp,
Prog.\ Theor.\ Phys.\ \textbf{105}, 1 (2001);
C.\ Honerkamp and M.\ Salmhofer,
Phys.\ Rev.\ B \textbf{64}, 184516 (2001).
%--------------------------
%--------------------------
\bibitem{Zanchi2001} 
D.\ Zanchi,
Europhys.\ Lett.\ \textbf{55}, 376 (2001).
%--------------------------
%--------------------------
\bibitem{Honerkamp2003}
C.\ Honerkamp and M.\ Salmhofer,
Phys.\ Rev.\ B \textbf{67}, 174504 (2003).
%--------------------------
%--------------------------
\bibitem{Katanin2004}
A.A.\ Katanin and A.P.\ Kampf,
Phys.\ Rev.\ Lett.\ \textbf{93}, 106406 (2004).
%--------------------------
\bibitem{Metzner}
D.\ Rohe and W.\ Metzner,
Phys.\ Rev.\ B \textbf{71}, 115116 (2005);
W.\ Metzner, J.\ Reiss, and D.\ Rohe,
Phys.\ Stat.\ Sol.\ B \textbf{243}, 46 (2006).
%--------------------------
%--------------------------
\bibitem{Kishine1999}
J.\ Kishine and K.\ Yonemitsu,
Phys.\ Rev.\ B \textbf{59}, 14823 (1999).
%--------------------------
%--------------------------
\bibitem{Freire}
H.\ Freire, E.\ Correa, and A.\ Ferraz,
Phys.\ Rev.\ B \textbf{71}, 165113 (2005).
%--------------------------
%------------------------
\bibitem{Bourbonnais_review}
For a review,
C.\ Bourbonnais and D.\ J\'erome, 
in \textit{Advances in Synthetic Metals, Twenty Years of Progress in Science
and Technology}, edited by P.\ Bernier, S.\ Lefrant, and
G.\ Bidan (Elsevier, New York, 1999), p. 206.
%--------------------------
%--------------------------
\bibitem{Kishine1998}
J.\ Kishine and K.\ Yonemitsu,
J.\ Phys.\ Soc.\ Jpn.\  \textbf{67}, 2590 (1998);
 \textbf{68}, 2790 (1999).
%--------------------------
%--------------------------
\bibitem{Lin1997}
H.H.\ Lin, L.\ Balents, and M.P.A.\ Fisher,
Phys.\ Rev.\ B \textbf{56}, 6569 (1997). 
%--------------------------
%--------------------------
\bibitem{Duprat2001}
R.\ Duprat and C.\ Bourbonnais,
Eur.\ Phys.\ J.\ B \textbf{21}, 219 (2001).
%--------------------------
%--------------------------
\bibitem{Bourbonnais2004}
C.\ Bourbonnais and R.\ Duprat,
J.\ Phys.\ IV France \textbf{114}, 3 (2004).
%--------------------------
\bibitem{Fuseya2005}
Y.\ Fuseya and Y.\ Suzumura, 
J.\ Phys.\ Soc.\ Jpn.\ \textbf{74}, 1263 (2005).
%--------------------------
\bibitem{Dupuis2005}
N.\ Dupuis, C.\ Bourbonnais, and J.C.\ Nickel,
cond-mat/0510544;
J.C.\ Nickel, R.\ Duprat, C.\ Bourbonnais, and N.\ Dupuis,
Phys.\ Rev.\ B \textbf{73}, 165126 (2006).
%------------------------
\bibitem{Fuseya2006}
Y.\ Fuseya, M. Tsuchiizu, Y.\ Suzumura, and C.\ Bourbonnais, 
preprint.
%--------------------------
%--------------------------
\bibitem{Doucot2003}
S.\ Dusuel and B.\ Dou\c{c}ot,
Phys.\ Rev.\ B \textbf{67}, 205111 (2003).
%--------------------------
%--------------------------
\bibitem{Giamarchi2001}
S.\ Biermann, A.\ Georges,\ A.\ Lichtenstein, and T.\ Giamarchi, 
Phys.\ Rev.\ Lett.\ \textbf{87}, 276405 (2001).
%--------------------------
%--------------------------
\bibitem{Giamarchi2002}
S.\ Biermann, A.\ Georges, T.\ Giamarchi, and A.\ Lichtenstein,
in \textit{Strongly Correlated Fermions and Bosons 
  in Low-Dimensional Disordered Systems},
 edited by I.V.\ Lerner, B.L.\ Althsuler, V.I.\ Fal'ko,
    and T.\ Giamarchi
  (Kluwer Academic Publishers, Netherlands, 2002), p.\ 81.
%--------------------------
%--------------------------
\bibitem{Giamarchi2004}
T.\ Giamarchi, S.\ Biermann, A.\ Georges, and A.\ Lichtenstein,
J.\ Phys.\ IV France \textbf{114}, 23 (2004).
%--------------------------
%--------------------------
\bibitem{Berthod}
C.\ Berthod, T.\ Giamarchi, S.\ Biermann, and A.\ Georges,
cond-mat/0602304.
%--------------------------
%--------------------------
\bibitem{Essler2002}
F.H.L.\ Essler and A.M.\ Tsvelik,
Phys.\ Rev.\ B \textbf{65}, 115117 (2002).
%--------------------------
%--------------------------
\bibitem{Dagotto}
For a review,
E.\ Dagotto and T.M.\ Rice,
 Science \textbf{271}, 618 (1996),
and references therein.
%--------------------------
%--------------------------
\bibitem{Fabrizio1993}
M.\ Fabrizio, 
Phys.\ Rev.\ B \textbf{48}, 15838 (1993).
%--------------------------
%--------------------------
\bibitem{Nersesyan1993}
A.A.\ Nersesyan, A.\ Luther, and F.V. Kusmartsev,
Phys.\ Lett.\ A \textbf{176}, 363 (1993).
%--------------------------
%--------------------------
\bibitem{Khveshchenko1994}
D.V.\ Khveshchenko and T.M.\ Rice,
Phys.\ Rev.\ B \textbf{50}, 252 (1994).
%--------------------------
%--------------------------
\bibitem{Schulz1996}
H.J.\ Schulz,
Phys.\ Rev.\ B \textbf{53}, R2959 (1996); in
\textit{Correlated Fermions and Transport in Mesoscopic Systems}, 
edited by T.\ Martin, G.\ Montambaux, and T.\ Tr\^an Thanh V\^an
(Editions Fronti\`eres, Gif-sur-Yvette, France, 1996), p.\ 81.  
%--------------------------
%--------------------------
\bibitem{Balents1996}
L.\ Balents and M.P.A.\ Fisher, 
Phys.\ Rev.\ B \textbf{53}, 12133 (1996). 
%--------------------------
%--------------------------
\bibitem{Lin1998}
H.H.\ Lin, L.\ Balents, and M.P.A.\ Fisher,
Phys.\ Rev.\ B \textbf{58}, 1794 (1998).
%--------------------------
%--------------------------
\bibitem{Tsuchiizu1999}
M.\ Tsuchiizu and Y.\ Suzumura,
Phys.\ Rev.\ B \textbf{59}, 12326 (1999).
%--------------------------
%--------------------------
\bibitem{Tsuchiizu2002}
M.\ Tsuchiizu and A.\ Furusaki,
  Phys.\ Rev.\ B \textbf{66}, 245106 (2002).
%--------------------------
%--------------------------
\bibitem{Fradkin2003}
C.\ Wu, W.V.\ Liu, and E.\ Fradkin,
  Phys.\ Rev.\ B \textbf{68}, 115104 (2003).
%--------------------------
%--------------------------
\bibitem{Tsuchiizu2005}
M.\ Tsuchiizu and Y.\ Suzumura,
Phys.\ Rev.\ B \textbf{72}, 075121 (2005).
%--------------------------
%--------------------------
\bibitem{Noack}
R.M.\ Noack, S.R.\ White, and D.J.\ Scalapino,
  Phys.\ Rev.\ Lett.\ \textbf{73}, 882 (1994);
  Physica C \textbf{270}, 281 (1996).
%--------------------------
%--------------------------
\bibitem{Weihong}
Z.\ Weihong, J.\ Oitmaa, C.J.\ Hamer, and R.J.\ Bursill,
J.\ Phys.: Condens.\ Matter \textbf{13}, 433 (2001).
%--------------------------
%--------------------------
\bibitem{Gogolin}
A.O.\ Gogolin, A.A.\ Nersesyan and A.M.\ Tsvelik,
  \textit{Bosonization and Strongly Correlated Systems}
  (Cambridge University Press, Cambridge, 1998).
%--------------------------
%--------------------------
\bibitem{Tsuchiizu_Furusaki}
M.\ Tsuchiizu and A.\ Furusaki,
  Phys.\ Rev.\ Lett.\ \textbf{88}, 056402 (2002);
  Phys.\ Rev.\ B \textbf{69}, 035103 (2004).
%--------------------------
%--------------------------
\bibitem{Yang}
C.N.\ Yang, 
Phys.\ Rev.\ Lett.\ \textbf{63}, 2144 (1989);
C.N.\ Yang and S.C.\ Zhang,
Mod.\ Phys.\ Lett.\ \textbf{4}, 759 (1990);
M.\ Pernici,
Europhys.\ Lett.\ \textbf{12}, 75 (1990);
S.C.\ Zhang,
Phys.\ Rev.\ Lett.\ \textbf{65}, 120 (1990);
H.J.\ Schulz, in
\textit{The Hubbard Model}, edited by D.\ Baeriswyl \textit{et al.}
  (Plenum, New York, 1995), p.\ 89.
%--------------------------
%--------------------------
\bibitem{Ovchinikov}
A.A.\ Ovchinikov,
Sov.\ Phys.\ JETP \textbf{30}, 1160 (1970).
%--------------------------
%--------------------------
\bibitem{Tsuchiizu_unpublished}
M.\ Tsuchiizu,  Y.\ Suzumura, and C.\ Bourbonnais,
unpublished.
%--------------------------
%--------------------------
\bibitem{Shelton}
D.G.\ Shelton, A.A.\ Nersesyan, and A.M.\ Tsvelik,
Phys.\ Rev.\ B \textbf{53}, 8521 (1996);
A.A.\ Nersesyan and A.M.\ Tsvelik,
Phys.\ Rev.\ Lett. \textbf{78}, 3939 (1997);
A.M.\ Tsvelik,
Phys.\ Rev.\ B \textbf{42}, 10499 (1990).
%--------------------------
%--------------------------
\bibitem{Shiba}
H.\ Shiba, 
Prog.\ Theor.\ Phys.\ \textbf{48}, 2171 (1972);
%--------------------------
%--------------------------
\bibitem{Nagaoka}
Y.\ Nagaoka,
Prog.\ Theor.\ Phys.\ \textbf{52}, 1716 (1974). 
%--------------------------


\end{thebibliography}
\end{document}